\documentclass[sigconf]{acmart}
\pdfoutput=1
\usepackage{graphicx}
\usepackage{subcaption}
\usepackage{booktabs}
\usepackage[noabbrev,capitalise]{cleveref}
\usepackage{flushend}
\usepackage{multicol, blindtext}
\setcopyright{none}
\renewcommand\footnotetextcopyrightpermission[1]{}
\acmDOI{}

\acmISBN{}

\acmConference[]{}{}{}
\acmPrice{}

\begin{document}
\title[Detection of paroxysmal atrial fibrillation using Bidirectional RNN]{Detection of Paroxysmal Atrial Fibrillation using Attention-based Bidirectional Recurrent Neural Networks}

\author{Supreeth P. Shashikumar}
\affiliation{%
  \department{School of Electrical and Computer Engineering}
  \institution{Georgia Institute of Technology}
  \city{Atlanta}
  \state{Georgia, USA}
}
\email{supreeth@gatech.edu}

\author{Amit J. Shah}
\affiliation{%
  \department{Department of Epidemiology}
  \institution{Rollins School of Public Health, Emory University}
  \city{Atlanta}
  \state{Georgia, USA}
}
\email{ajshah@emory.edu}

\author{Gari D. Clifford}
\affiliation{%
  \department{Department of Biomedical Informatics}
  \institution{Emory University}
  \department{Department of Biomedical Engineering}
  \institution{Georgia Institute of Technology}
  \city{Atlanta}
  \state{Georgia, USA}
  }
\email{gari@gatech.edu}

\author{Shamim Nemati}
\affiliation{%
  \department{Department of Biomedical Informatics}
  \institution{Emory University}
  \city{Atlanta}
  \state{Georgia, USA}
  }
\email{shamim.nemati@emory.edu}

%FROM AMIT: I would not capitalize "atrial fibrillation" or "paroxysmal" - also would be good to say a few words to describe your cohort like "ICU patients" 
\begin{abstract}
Detection of atrial fibrillation (AF), a type of cardiac arrhythmia, is difficult since many cases of AF are usually clinically silent and undiagnosed. In particular paroxysmal AF is a form of AF that occurs occasionally, and has a higher probability of being undetected. In this work, we present an attention based deep learning framework for detection of paroxysmal AF episodes from a sequence of windows. Time-frequency representation of 30 seconds recording windows, over a 10 minute data segment, are fed sequentially into a deep convolutional neural network for image-based feature extraction, which are then presented to a bidirectional recurrent neural network with an attention layer for AF detection. To demonstrate the effectiveness of the proposed framework for transient AF detection, we use a database of 24 hour Holter Electrocardiogram (ECG) recordings acquired from 2850 patients at the University of Virginia heart station. The algorithm achieves an AUC of 0.94 on the testing set, which exceeds the performance of baseline models. We also demonstrate the cross-domain generalizablity of the approach by adapting the learned model parameters from one recording modality (ECG) to another (photoplethysmogram) with improved AF detection performance. The proposed high accuracy, low false alarm algorithm for detecting paroxysmal AF has potential applications in long-term monitoring using wearable sensors.
\end{abstract}

\renewcommand{\shortauthors}{S.P Shashikumar et al.}

\maketitle

\section{Introduction}
Atrial fibrillation (AF), an arrhythmia that arises from irregular atrial contraction, has a prevalence of 2\% in the adult population \cite{camm20122012,ball2013atrial}. AF can often lead to dangerous complications such as ischemic stroke and heart failure \cite{stewart2002population,wong2012increasing}, but these complications can be treated or can be avoided by anticoagulation and control of heart rate or rhythm \cite{fuster2006acc}. This highlights the need for developing accurate AF detection methods, which would allow for early treatment and diagnosis of AF.

The difficulty in the diagnosis of AF stems from its asymptomatic and paroxysmal nature. Moreover, when it is eventually diagnosed, clinicians must make multifactorial decisions regarding the optimal course of therapeutic interventions, which, among other factors, may require monitoring the burden of AF (defined as the amount of time spent in AF). Clinically, AF is diagnosed by the absence of P waves in the Electrocardiogram (ECG) with a rapid, irregular rhythm. However, clinical adjudication based on ECG alone is difficult in the presence of recording noise and distortion, which is typical in ambulatory monitoring. Over the last four decades, numerous methods have been proposed in the literature for AF detection based on beat-to-beat timing and morphological and frequency domain properties of the ECG. 
These range from hidden Markov modeling of the beat-to-beat timing \cite{moody1983new}, to QRST subtraction and frequency analysis of the residual waveform \cite{stridh2001spatiotemporal}, as well as many other threshold-based, heuristic and machine learning approaches \cite{tateno2001automatic,lake2010accurate,linker2009long,colloca2013support,petrenas2012echo}. More recently, we have developed deep learning approaches applied to cardiac signals recorded in real world noisy environments \cite{shashikumar2017deep} in an attempt to discover novel patterns indicative of AF. Similar deep learning approaches were subsequently reported (along with many other standard approaches) in the recent PhysioNet Challenge, an international competition focused on the classification of over 10,000 publicly available ECGs \cite{clifford2017af}. This competition, organized over 9 months, led to 45 publications and 75 independent pieces of software, most of which were open sourced \cite{Physionet2017Site}. Notably, this competition provided the first slew of deep learning based ECG rhythm detectors, along with some simultaneous pre-prints. However, all these methods were developed within a standard machine learning paradigm, ignoring the long tracts of non-arrhythmic data that often exist between sporadic episodes of arrhythmia. This leads to wholly optimistic estimates of the performances of these algorithms when applied to long term recordings of several hours or days, and significantly underestimates the false alarm rate. The work presented here addresses the issue of false alarms through the use of an attention model applied to a deep learning framework.

Our goal during the development of the algorithm was three fold: 
\begin{itemize}
\item To assess whether feature vectors extracted from the deep learning pipeline in combination with time series covariates, extracted using traditional methods and hand-crafted features, can achieve improved performance in comparison to other ECG based approaches. 
\item To test whether employing a soft attention mechanism would improve the performance of the algorithm in the presence of paroxysmal AF, by learning to put more weight on the windows with higher potential prevalence of AF, as opposed to noise and other types of rhythm irregularity.
\item To empirically evaluate the potential of the model for transfer learning across different source domains.
\end{itemize}
To achieve these objectives, we designed a deep learning model based on a Bidirectional Recurrent Neural Network (BRNN) for detection of AF from a three lead ECG. The proposed model extracts spectral features from ECG signal using wavelet analysis, and a convolutional neural network (CNN) is used to process these spectral features, followed by a BRNN to analyze these features in temporal order to detect AF from 30 second windows spanned over 10 minutes of recording. In addition, we also implement a soft attention mechanism on top of the BRNN to enable the algorithm to prioritize ECG segments that are predictive of AF (given the paroxysmal nature of AF). The entire pipeline of the proposed model is illustrated in \cref{fig:block_diag}. 

\section{Data}
\subsection{Holter ECG dataset}
The dataset consists of 24 hour Holter ECG recordings collected from 2850 patients at the University of Virginia (UVA) Heart Station from 12/2004 to 10/2010. The age of the patients varied from a few months to 100 years, with an average  ($\pm$ standard deviation) value of 47 $\pm$ 25 years. AF labels were obtained from an automatic classifier (Philips Holter Software), and each record was examined by a clinical adjudicator to confirm presence or absence of AF. The dataset was found to consist of various heart rhythms including AF, Normal Sinus Rhythm (NSR), and Sinus Rhythm with ectopy. After excluding segments with a low signal quality index ($SQI$) \cite{li2007robust}, which accounted for less than 2\% of all segments, the remaining recordings were divided into 364,012 10 minute segments. Each such segment was classified as \emph{AF} if the burden of AF was greater than 5\% (i.e. for more than 30 seconds), and \emph{Other Rhythm} otherwise, however for completeness we also provide results for different burdens of AF. Although, labels obtained from automatic systems can be noisy at the level of beat-by-beat annotation, our goal was to quantify presence of AF over consecutive 10 minute segments, which is less prone to noise due to averaging of labels and removal of low $SQI$ segments. Moreover, we utilize ECG-based AF detection as a starting point for cross-domain transfer learning to demonstrate the utility of using external data and data types. In particular, we use a database of pulsatile photoplethysmogram (PPG) recordings of AF from a smart watch, as described next, for this purpose. 

\subsection{Smart Watch PPG dataset}\label{subsection:simband}
All subjects were adult patients (18-89 years old) who were hospitalized and were undergoing telemetry monitoring at the Emory University Hospital (EUH),  Emory University Hospital Midtown (EUHM), and Grady Memorial Hospital (GMH). The study was approved by the institutional review board (IRB \# 00084629) of the Emory University Hospital and Grady Memorial Hospital in Atlanta, GA. Patients were recruited at random; the rhythms were reviewed by an ECG technician, physician study coordinator, and cardiologist. The study took place from October 2015 through March 2016. Using a modern research watch-based wearable device (the Samsung Simband \cite{simbandSite}), we recorded ambulatory PPG data from $97$ subjects, $44$ with \emph{AF} and $53$ with \emph{other rhythms} for approximately 5-10 minutes. Simultaneous 128 Hz data were captured form a single channel ECG, multi-wavelength ($8$ channel) PPG and a tri-axial accelerometer.

\begin{figure}[t]
\centering
\includegraphics[width=8cm, height = 7cm]{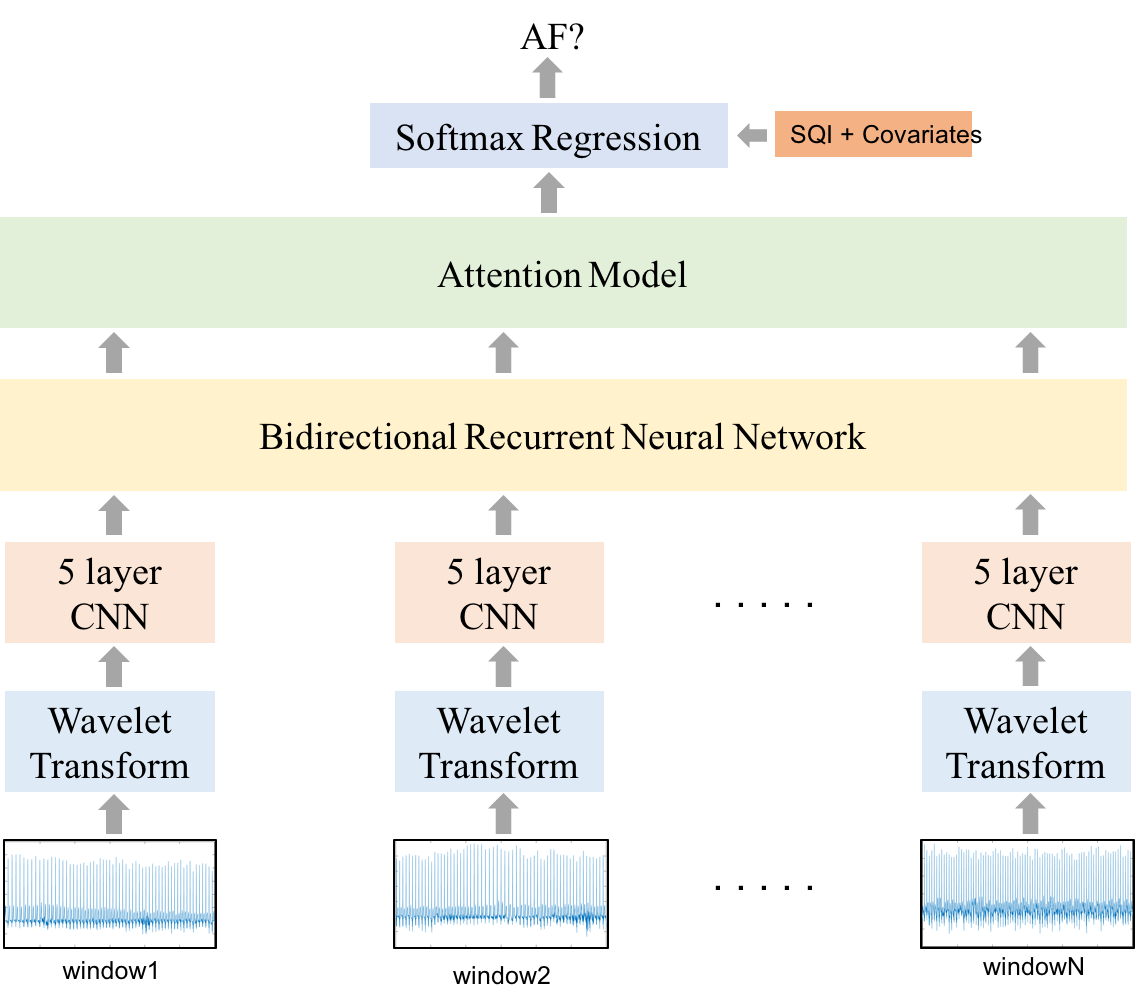} 
  \caption{Schematic diagram of the AF detection algorithm. Each 10 minute ECG segment was split into 30 second windows, and a CNN was used to extract time-frequency features from the spectrograms obtained from each of the 30 second windows. The output feature vectors from the CNN were then fed sequentially into a BRNN, a soft attention mechanism prioritized the segments of ECG to attend, and then a softmax regression layer detected AF in the ECG segment by combining the deep learning features with other time series covariates.}
  \label{fig:block_diag}
\end{figure}
\section{Methods}
\subsection{Model Architecture} \label{subsection:model-arch}

%The proposed algorithm was designed to detect the presence of AF in longitudinal ECG recordings, by first dividing the records into 10 minute segments, and then breaking each segment into 30 second windows. Our objective was to detect the presence of AF and produce a label once every 10min. 
An overview of the proposed AF detector is shown in \cref{fig:block_diag}. The algorithm included the following stages:
\begin{itemize}
\item \emph{Data preparation}: Splitting longitudinal recordings into non-overlapping 10 minute multi-channel ECG segments sampled at 200Hz.
\item \emph{Pre-processing}: Filtering to reduce and remove noise from the ECG signals, and selecting the channel with the highest $SQI$. Further split the 10 minute segment to consecutive 30 second windows.
\item \emph{Time series covariates}: Computing covariates based on beat-to-beat interval variations from the 10 minute segments. 
\item \emph{Frequency analysis}: extracting time-frequency information from the processed ECG signals by applying a wavelet transform.
\item \emph{Feature extraction using CNN}: Feeding the wavelet spectrograms into a CNN for feature extraction.
\item \emph{Bidirectional Recurrent Neural Network}: Feeding features extracted from the CNN into a BRNN to capture temporal patterns.
\item \emph{Attention layer}: Passing the data from the previous step through a soft attention mechanism that assigns more weights to the windows with higher prevalence of AF.
\item \emph{Classification layer}: Feeding the weighted feature vector from the attention layer and the time series covariates into a fully connected layer. The classification layer then computes the likelihood of presence of AF in the input ECG segment.
\end{itemize}

\subsection{Data preparation and pre-processing}
Each ECG channel was lowpass filtered to remove frequencies above 40Hz, using a Butterworth filter of order five. Next, QRS (beat) detection was performed on each channel individually using two QRS detectors - $jQRS$ \cite{behar2013ecg,johnson2015multimodal,johnson2014r} and $waveletQRS$ \cite{martinez2004wavelet} and the resultant beats were used to compute the signal quality index $bSQI$ \cite{li2007robust}. To remove the noisy segments, we excluded all  segments with $bSQI$ < 0.90, which eliminated roughly 2\% of the data segments. In the remaining segments, the channel with the highest $bSQI$ was then chosen for further analysis.

\subsection{Time series covariates}
We incorporated several metrics based on beat-to-beat interval variations in our AF detection algorithm. We calculated
the standard deviation of the beat-to-beat interval time series
($STD$), as well as a robust version of the standard deviation
($STD_r$), after discarding the intervals outside the 0.02-0.98
percentile range (we assumed that the extreme intervals
were due to erroneous pulse onset detection). We also calculated the sample entropy \cite{richman2000physiological} of the beat-to-beat interval time series with the embedding dimensions of 
$m$ = 1, and 2
(denoted $SampEn_1$ and $SampEn_2$, respectively).

\subsection{Wavelet decomposition}
\begin{figure}[h]
\centering
\includegraphics[width=8cm, height = 7cm]{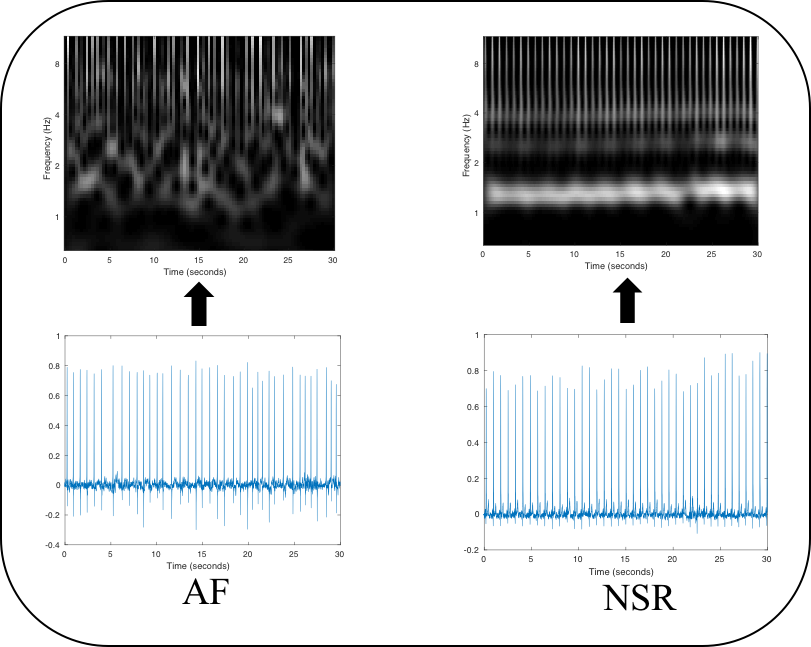} 
  \caption{Wavelet power spectrum for 30 second ECG sample. The left panel shows an example of AF segment and the right panel represents a Normal Sinus Rhythm segment}
  \label{fig:spectrogram}
\end{figure}
Wavelet transform is an ideal tool for extracting time-frequency information from signals of a non-stationary nature, due to its optimal time-frequency resolution trade-offs \cite{daubechies1990wavelet}.
AF is characterized by irregular heart beat rhythms \cite{fuster2006acc}, which can be captured through the spectral analysis of the ECG signal (See \cref{fig:spectrogram}). \citet{shashikumar2017deep} proposed an AF detection approach that was based on employing wavelet decomposition to extract features from the frequency domain. In our algorithm, the 10 minute segment was split into non-overlapping 30 second windows, and the wavelet transform \cite{shashikumar2017deep} was applied to each of the windows, resulting in a wavelet spectrogram matrix of size 20 by 300. The hyper-parameters used during the wavelet decomposition were as follows: maximum scale of 20; 1/15 octaves per scale; minimum scale of 8.3; and a down-sampling factor of 20 along the time axis. The mathematics behind the wavelet analysis is well documented in \citet{grinsted2004application}. In practice, recording noise and artifacts might corrupt the spectral content present in the signal which makes it difficult to distinguish AF rhythm from the other rhythms. To reduce the effect of the recording noise and other artifacts, we thresholded the spectrograms using surrogate data analysis as described in \citet{shashikumar2017deep}. 

\subsection{Convolutional neural network for feature extraction}
\begin{figure}[h]
\centering
\includegraphics[width=8cm, height = 3cm]{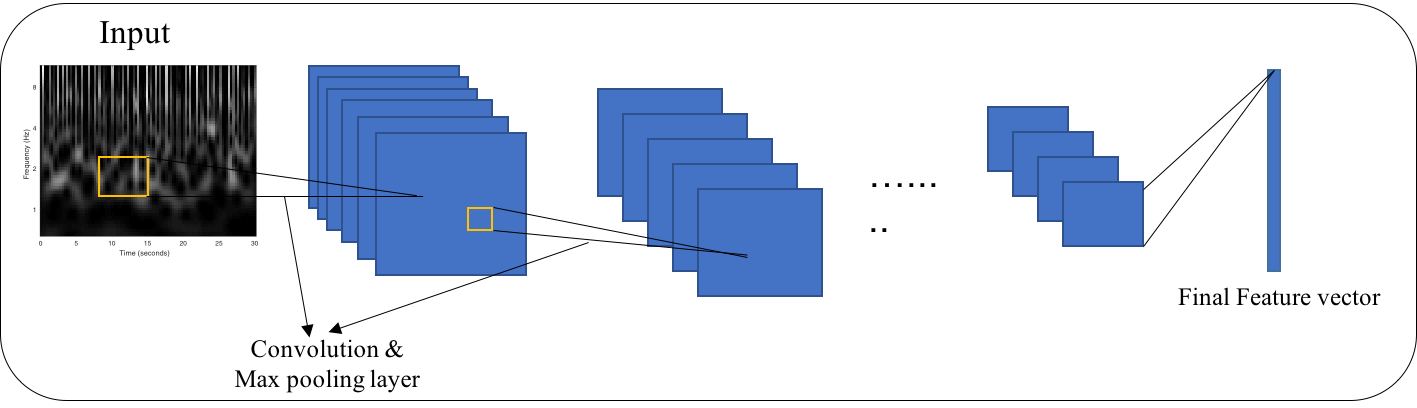} 
  \caption{Schematic diagram of the deep CNN layer. The spectrogram obtained from the 30 second ECG window is fed into a 5 layer CNN and the output feature vector is then sequentially fed into BRNN}
  \label{fig:cnn}
\end{figure}
CNNs have been shown to be very effective in image-based classification tasks \cite{lecun1995convolutional,krizhevsky2012imagenet} due to their  translation-invariance properties, ability to capture local features, shared-weights architecture that enhances generalizability \cite{lecun1998gradient}.
 Time-frequency representation of pulsatile cardiac timeseries of NSR and AF tend to exhibit distinct patterns (see \cref{fig:spectrogram}), although the exact location of the dominant frequency depends on the underlying heart rate. The translation-invariance property of the CNN can handle such spectral shifts in the patterns of interest. Our CNN architecture, comprised of two successive convolutional layers (each layer having a kernel size of 3x21) and a max pooling layer, followed by two more convolutional layers (each having a kernel size of 4x21), and one fully connected layer. Each convolutional layer in the architecture was followed by a layer of activation with Rectified Linear Unit (ReLU) nonlinearity. All the pooling layers had pooling region of size 2x2 with a stride of 2 along both the directions. The number of filters used for each of the convolutional layers was 10, with the fully connected layer having a total of 50 filters. The output of the CNN was a 50 dimensional feature vector that summarized the entire wavelet power spectrum. 

\subsection{Bidirectional recurrent neural network}
\begin{figure}[h]
\centering
\includegraphics[width=8cm, height = 3cm]{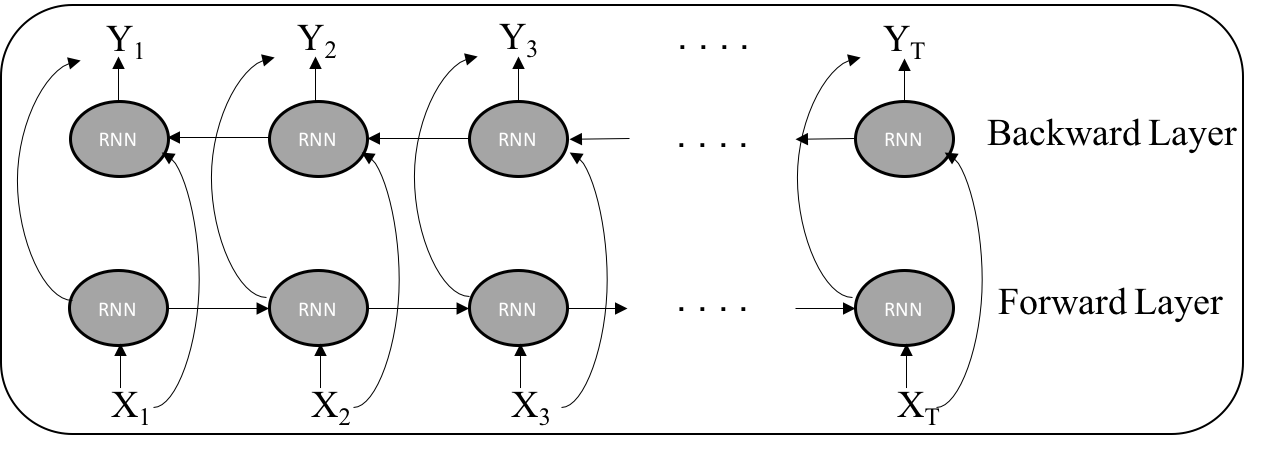} 
  \caption{Schematic diagram of the bidrectional recurrent neural network}
  \label{fig:birnn}
\end{figure}
Recurrent neural networks (RNNs) are capable of capturing temporal patterns in the data \cite{lee1994dynamic,graves2013speech}. Standard RNNs are unidirectional, in the sense that input data is processed in temporal order. A shortcoming of this approach is that the RNNs are restricted to the use of previous context. 
%In our problem, the entire 10 minute segment is available at once which motivated us to employ an architecture that makes use of past and future contexts. 
BRNNs \cite{schuster1997bidirectional} provide a solution by processing the data in both forward and backward directions. \Cref{fig:birnn} shows a BRNN architecture unfolded in time for $T$ time steps. The BRNN consists of a forward layer and a backward layer. The forward  layer $h_t^f$ is computed by processing the input data from $t = 1, ..., T $, and the backward layer $h_t^b$ is computed by processing the input data from $t = T, ..., 1$ with output from both layers combined as follows:
\begin{equation}
h_t^f = tanh(W_{xh}^fx_t + W_{hh}^fh_{t-1} + b_h^f)
\end{equation}
\begin{equation}
h_t^b = tanh(W_{xh}^bx_t + W_{hh}^bh_{t+1} + b_h^b)
\end{equation}
\begin{equation}
y_t = W_{hy}^fh_t^f + W_{hy}^bh_t^b + b_y)
\end{equation}

In our model, the size of the hidden state in the forward and the backward layer was 50, with the output at each time step being a 100 dimensional vector. The output vectors $y_t$ obtained by processing the sequence of input data from $t = 1, ..., T$ was then fed into an attention layer.

\subsection{Attention layer} \label{subsection:attention-layer}
Paroxysmal atrial fibrillation (PAF) occurs as intermittent periods of AF scattered with episodes of normal sinus rhythm \cite{lip2001paroxysmal}. Thus, a soft attention mechanism \cite{bahdanau2014neural,zhou2016attention,yang2016hierarchical} was employed on top of the BRNN so that more emphasis (or attention) could be put on windows with higher prevalence of AF. The attention mechanism can be formulated as follows - Let the outputs from the BRNN [{$y_1, y_2, ...., y_T$}] be combined into a matrix \emph{Y} which is of size $N$ x $T$, where $T$ is the length of the input sequence and $N$ is the size of output vector $y_t$. The weighted output vector $h_{att}$ of the attention layer is as follows - 
\begin{equation}
\alpha = softmax(w_{att}^TY)
\end{equation}
\begin{equation}
h_{att} = Y\alpha^T
\end{equation}

That is, $\alpha$ is a weight vector computed from matrix $Y$, and the output $h_{att}$ calculated as a weighted sum of all the output vectors from the BRNN ($h_{att}$ was a 100 dimensional vector).

\subsection{Classification layer}
The attention layer was then followed by a fully connected layer, the output of which was fed into a softmax regression layer. In addition to the features obtained from the BRNN, time series covariates were fed as inputs to the softmax regression layer. The output from the softmax layer corresponded to the likelihood of presence of AF in the input ECG segment.
 
\section{Results}
\subsection{Experimental setup}
Of the 2850 patients, 80\% of them were used for developing the model (training set) and the remaining 20\% of the patients were used as the hold out test set. The training set consisted of a total of 2290 patients out of which 217 patients contained at least one episode of AF, and the testing set consisted of a total of 575 patients out of which 55 patients contained at least one episode of AF (we refer to an episode of AF as a 10 minute segment with >5\% of AF burden). Splitting the recordings of each patient to 10 minute segments resulted in a total of 291,240 segments (21,542 AF segments) in the training set and 72,772 segments (5,419 AF segments) in the testing set. 

Details regarding the training and development of the model are as follows: The batch size was fixed at 90 patients (70\% \emph{Other Rhythm} patients, 30\% \emph{AF} patients), with data randomly sampled (with replacement) in every epoch. The model was trained end-to-end for a total of 200 epochs. RMSProp \cite{tieleman2012lecture} was used as the optimizer and L2 regularizer with regularization parameter $\lambda$ = 0.01 was used. The learning rate was set at  0.001. All the hyper-parameters of the model: number of filters in each of the CNN layer, size of the hidden layer in the BRNN, learning rate, regularization parameter $\lambda$ were optimized using Bayesian Optimization technique \cite{ghassemi2014global}. A subset of the training set was used for hyper-parameter optimization.
Area under the receiver operating characteristic ($AUC$) curve, Area under the precision recall ($AUC_{pr}$) curve, accuracy, specificity were calculated for both the training and the test sets. The sensitivity level was fixed at 0.85. All pre-processing of the data, time series covariates computation and wavelet decomposition were performed in Matlab \cite{MATLAB2016}. The rest of the pipeline was implemented in python with CNN and BRNN implemented using Tensorflow \cite{Tensorflow}.

\subsection{AF detection performance}
First, a comparative analysis was done to evaluate the performance of the model based on \% of AF burden. The \% of AF burden determines the presence of AF in the 10 minute segments. The performance of the model with various thresholds on the AF burden are shown in \cref{table:afburden}. It can be seen that labeling segments as AF with AF burden $>$5\% resulted in an AUC of 0.94 on the testing set. It should be noted that with increasing threshold on the AF burden, a majority of the segments containing PAF will be filtered out. The model trained on segments with $>$75\% AF burden performed the best in comparison to the others, likely due to the absence of PAF segments (which are more difficult to detect). Since our goal was to detect both PAF and AF segments, we labeled segments with AF burden $>$5\% as AF in all the following experiments.

\begin{table}[t]
\centering
\caption{Summary of classifier performance for different \% of AF burden. The Area Under the Curve ($AUC$), Area under the precision recall curve ($AUC_{pr}$), Specificity ($SPC$) and Accuracy ($ACC$) are reported for both training set and testing set}\vspace{-.2cm}
\label{table:afburden}
\resizebox{\columnwidth}{!}{
\begin{tabular}{cc c c c c c c c}
  \toprule
  & \multicolumn{4}{c}{Testing set} & \multicolumn{4}{c}{Training set} \\
  \cmidrule(l){2-5} \cmidrule(l){6-9}
  
  \% AF burden & $AUC$ & $AUC_{pr}$ & $SPC$ & ACC & $AUC$ & $AUC_{pr}$ & $SPC$ & ACC \\
  
  \cmidrule(l){1-1} \cmidrule(l){2-2} \cmidrule(l){3-3} \cmidrule(l){4-4} \cmidrule(l){5-5} \cmidrule(l){6-6} \cmidrule(l){7-7} \cmidrule(l){8-8} \cmidrule(l){9-9} 
  
  $>$5 					& 0.94 & 0.84 & 0.95 & 0.94 & 0.96 & 0.93 & 0.96 & 0.96 \\
  $>$25 				& 0.93 & 0.82 & 0.92 & 0.92	& 0.96 & 0.93 & 0.95 & 0.95  \\
  $>$50 				& 0.95 & 0.82 & 0.93 & 0.95 & 0.97 & 0.91 & 0.95 & 0.94 \\
  $>$75 				& 0.97 & 0.84 & 0.96 & 0.96 & 0.98 & 0.96 & 0.98 & 0.96 \\
  \bottomrule
\end{tabular}
}
\end{table}

Conventionally, when RNNs are used in classification tasks the hidden state vector from the last time step alone is used for classification (no pooling). An alternative would be to compute the mean of all the hidden state vectors across all time steps, which is then used for classification (Mean pooling). A third alternative would be to use attention pooling which is described in \cref{subsection:attention-layer}. The performance of the model for the three kinds of pooling is shown in \cref{table:pooling} . It can be observed that using an attention layer (AUC of 0.94 on the testing set) provided improvement in performance over the other two pooling methods. An interesting observation to note is that mean pooling (AUC of 0.91 on the testing set) had a lower performance compared to performing no pooling (AUC of 0.92 on the testing set) at all.

\begin{table}[h]
\centering
\caption{Summary of classifier performance for different types of pooling. The Area Under the Curve ($AUC$), Area under the precision recall curve ($AUC_{pr}$), Specificity ($SPC$) and Accuracy ($ACC$) are reported for both training set and testing set} \vspace{-.2cm}
\label{table:pooling}
\resizebox{\columnwidth}{!}{
\begin{tabular}{l c c c c c c c c}
  \toprule
  & \multicolumn{4}{c}{Testing set} & \multicolumn{4}{c}{Training set} \\
  \cmidrule(l){2-5} \cmidrule(l){6-9}
  
  Model & $AUC$ & $AUC_{pr}$ & $SPC$ & ACC & $AUC$ & $AUC_{pr}$ & $SPC$ & ACC \\
  
  \cmidrule(l){1-1} \cmidrule(l){2-2} \cmidrule(l){3-3} \cmidrule(l){4-4} \cmidrule(l){5-5} \cmidrule(l){6-6} \cmidrule(l){7-7} \cmidrule(l){8-8} \cmidrule(l){9-9} 
  
  Attention pooling 			& \bf 0.94 & 0.84 & 0.95 & 0.94 & \bf 0.96 & 0.93 & 0.96 & 0.96 \\
  Mean pooling 					& 0.91 & 0.82 & 0.93 & 0.92 & 0.95 & 0.89 & 0.92 & 0.92 \\
  No pooling 					& 0.92 & 0.83 & 0.93 & 0.93 & 0.97 & 0.92 & 0.94 & 0.94 \\
  \bottomrule
\end{tabular}
}
\end{table}

\begin{figure*}[ht]
    \centering
    \begin{subfigure}[t]{0.3\textwidth}
        \centering
        \includegraphics[height = 2.3in, width=2.3in]{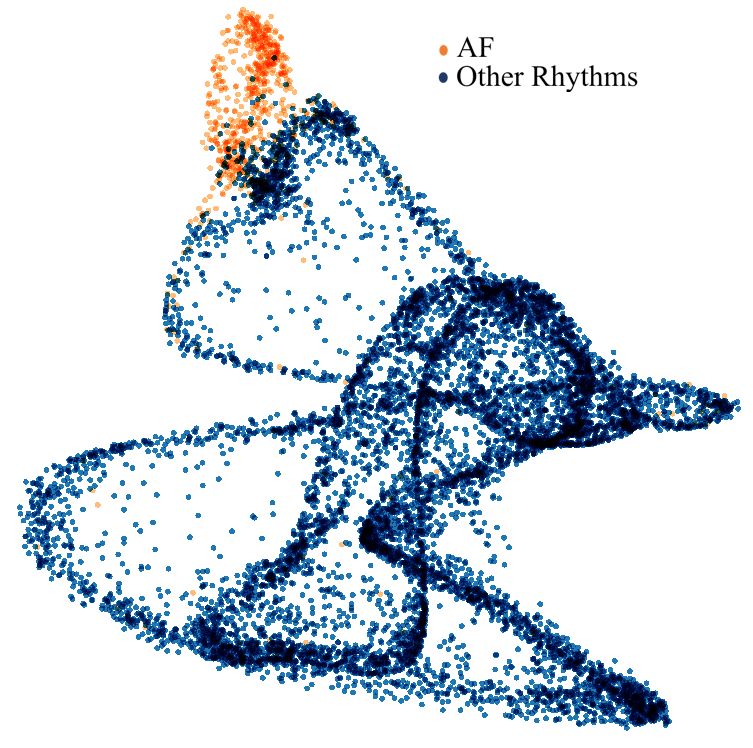}
        \caption{Visualization of data belonging to \emph{AF} and \emph{Other Rhythm} class}
        \label{subfig:overall}
    \end{subfigure}%
    \hfill
    \begin{subfigure}[t]{0.3\textwidth}
        \centering
        \includegraphics[height = 2.3in, width =2.3in]{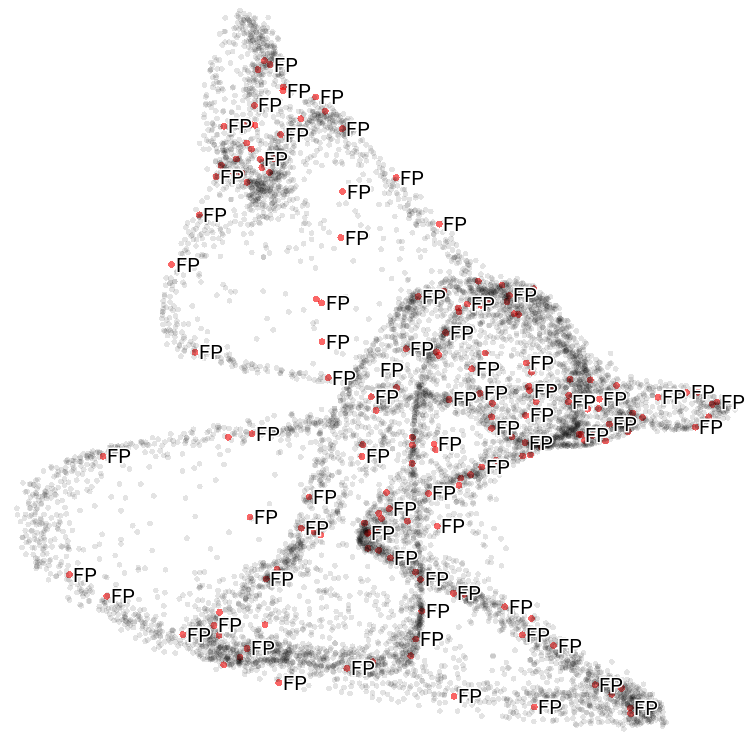}
        \caption{The False Positive (FP) cases highlighted}
        \label{subfig:FP}
    \end{subfigure}
    \hfill
    \begin{subfigure}[t]{0.3\textwidth}
        \centering
        \includegraphics[height = 2.3in, width =2.3in]{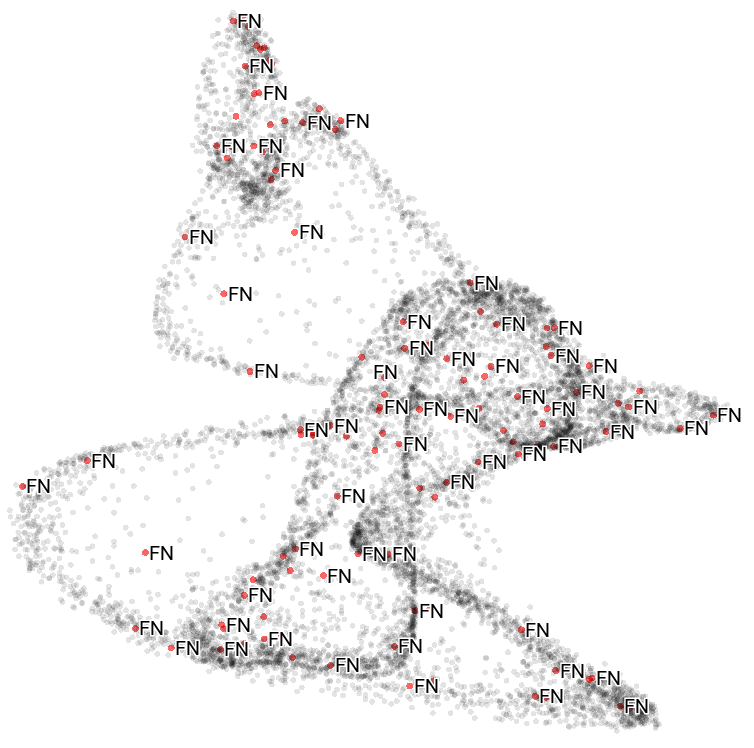}
        \caption{The False Negative (FN) cases highlighted}
        \label{subfig:FN}
    \end{subfigure}
    \caption{Visualization of the features extracted from the deep learning pipeline and the time series covariates by performing t-SNE \cite{maaten2008visualizing} based clustering. Clustering was performed on the testing dataset}
    \label{fig:embeddings}
\end{figure*}

Finally, a comparative study was performed to assess the importance of Spectrogram based features and time series covariates towards detection of AF. Three separate models were trained and tested - a) Using only the features extracted from the spectrogram and feeding them into the BRNN, b) Using only the time series covariates and feeding to the softmax regression layer (baseline model) and c) Combining both the spectrogram based features and time series covariates. In addition to the above models, we also compared the performance of our model with the AF detection algorithm proposed by \citet{carrara2015heart}. As proposed in \citet{carrara2015heart}, the Coefficient of Sample Entropy, the average Detrended Fluctuation Analysis value and the Local Dynamics score were computed for every 10 minute segments in the dataset, and were fed to a softmax regression layer for AF detection.  The results for the above outlined experiments have been tabulated in \cref{table:features}. The baseline model had the lowest performance (AUC of 0.87 on the testing set) as compared to the other two models. It can be observed that by combining the spectrogram based features and the time series covariates, the model achieved an AUC of 0.94 on the testing set.

\begin{table}[h]
\centering
\caption{Summary of classifier performance for three feature groups, or models. The Area Under the Curve ($AUC$), Area under the precision recall curve ($AUC_{pr}$), Specificity ($SPC$) and Accuracy ($ACC$) are reported for both training set and testing set}
\label{table:initialization}\vspace{-.2cm}
\resizebox{\columnwidth}{!}{
\begin{tabular}{l c c c c c c c c}
  \toprule
  & \multicolumn{4}{c}{Testing set} & \multicolumn{4}{c}{Training set} \\
  \cmidrule(l){2-5} \cmidrule(l){6-9}
  
  Model & $AUC$ & $AUC_{pr}$ & $SPC$ & ACC & $AUC$ & $AUC_{pr}$ & $SPC$ & ACC \\
  
  \cmidrule(l){1-1} \cmidrule(l){2-2} \cmidrule(l){3-3} \cmidrule(l){4-4} \cmidrule(l){5-5} \cmidrule(l){6-6} \cmidrule(l){7-7} \cmidrule(l){8-8} \cmidrule(l){9-9} 
  
   Spectrogram 				& 0.92 & 0.80 & 0.92 & 0.92 & 0.94 & 0.92 & 0.93 & 0.93 \\
   Covariates (Baseline) 	& 0.87 & 0.67 & 0.77 & 0.78 & 0.89 & 0.80 & 0.80 & 0.81 \\
  Combined 					& \bf 0.94 & 0.84 & 0.95 & 0.94 & \bf 0.96 & 0.93 &  0.96 & 0.96 \\
  \citet{carrara2015heart}  & 0.91 & 0.80 & 0.91 & 0.91 & 0.93 & 0.90 & 0.94 & 0.93 \\
  \bottomrule
\end{tabular}
}
\end{table}

\subsection{Visualization of the extracted features}
To understand the reason for the model to mis-classify the input ECG segments, we performed clustering on the features extracted from the deep learning pipeline and the time series covariates over 10 minute segments. We used t-SNE to perform the clustering \cite{maaten2008visualizing}. The visualization of the clusters obtained on the testing data is shown in \cref{subfig:overall}. Each point in the plot corresponds to a two-dimensional representation of features extracted from a 10 minute segment. The False Positive (FP) cases (i.e. \emph{Other Rhythm} classified as \emph{AF}) have been highlighted in \cref{subfig:FP}. The False Negative (FN) cases (i.e. \emph{AF} classified as \emph{Other Rhythm}) have been highlighted in \cref{subfig:FN}. Although, the two-dimensional embedding of the features only provides an approximate picture of the position of FP and FN cases, it sheds light on where the algorithm is most likely to disagree with the Holter monitor labels, which are known to be noise-prone \cite{drew2014insights}. Such points could make for potential candidates for active learning \cite{settles2012active}, wherein a meta-learning algorithm can iteratively refine the labels by querying an oracle (one or more clinicians) for ground truth labels, and re-train the model accordingly.

\subsection{Knowledge transfer across different source domains}
In recent times, the success of deep learning has been largely in part due to the availability of extensively labeled datasets \cite{wu2016google,silver2016mastering,mnih2015human,he2016deep}. Deep learning models are known to perform better when large sets of data is available. However, in practical scenarios especially in health care, access to such data is limited or not practical due to the amount of effort required. Transfer learning with deep neural networks provides a powerful framework to deal with such situations \cite{he2016deep,long2016deep,carlucci2017autodial,sun2016return,hsu2017learning}. In particular, we were interested in exploring how to transfer knowledge between domains, when the distribution of the input features and labels change, but the task (in our case, AF detection) remain the same. Given two datasets, one containing ECG recordings from 2850 patients and the other containing PPG recordings from only 97 patients, we aimed to transfer the knowledge learned from the ECG recordings (source domain) to PPG recordings (target domain) for detecting AF. 

The target domain in our experiment was a dataset of PPG signals recorded using Samsung Simband watches (described in \cref{subsection:simband}). We followed the same procedure as described in \cref{subsection:model-arch} for the wavelet decomposition and extraction of time series covariates from the PPG signal. We randomly split the patients as 80\% for training and the remaining 20\% for testing. We chose the same architecture that was used for the Holter ECG dataset, and we performed two experiments: a) We initialized the model with random weights and trained only on the Smart watch PPG dataset b) We initialized the variables in the model with the values learned from training on the Holter ECG dataset (Pre-trained model), and further refined the values with the Smart watch PPG training dataset. 

\begin{table}[h]
\centering
\caption{Comparison of classifier performance depending on the type of initialization of weights in the model for the PPG dataset. The Area Under the Curve ($AUC$), Area under the precision recall curve ($AUC_{pr}$), Specificity ($SPC$) and Accuracy ($ACC$) are reported for both training set and testing set} \vspace{-.2cm}
\label{table:features}
\resizebox{0.5\textwidth}{!}{
\begin{tabular}{l c c c c c c c c}
  \toprule
  & \multicolumn{4}{c}{Testing set} & \multicolumn{4}{c}{Training set} \\
  \cmidrule(l){2-5} \cmidrule(l){6-9}
  
  Type of initialization & $AUC$ & $AUC_{pr}$ & $SPC$ & ACC & $AUC$ & $AUC_{pr}$ & $SPC$ & ACC \\
  
  \cmidrule(l){1-1} \cmidrule(l){2-2} \cmidrule(l){3-3} \cmidrule(l){4-4} \cmidrule(l){5-5} \cmidrule(l){6-6} \cmidrule(l){7-7} \cmidrule(l){8-8} \cmidrule(l){9-9} 
  
   Random 				& 0.94 & 0.93 & 0.81 & 0.85 & 0.94 & 0.93 & 0.90 & 0.88 \\
   Pre-trained model 	& \bf 0.97 & 0.97 & 1.0 & 0.95  & \bf 0.94 & 0.92 & 0.90 & 0.89 \\
  \bottomrule
\end{tabular}
}
\end{table}

The improvement in performance obtained by using the pre-trained weights on Holter ECG dataset is shown in \cref{table:initialization}. We observe that when the model was \textbf{learned from scratch} (Model1), the model achieved an AUC of 0.94 on the testing set. Whereas when the model was \textbf{fine tuned from a pre-trained model} (Model2), the model achieved an AUC of 0.97 on the testing set. Comparing the testing set predictions of both the models, the p-value of comparing AUC curves \cite{cook2009use} was 0.0075, indicating a significant improvement. We achieved two important goals from these experiments - \textbf{First}, we have shown that the model that was initially trained on ECG data could be fine tuned to detect AF in PPG data. Even though the two datasets were based on completely different types of input signal, the deep learning based model was able to learn features that was generalizable across both the domains and which were predictive of AF. \textbf{Second}, we have shown that in scenarios where limited data is available, we can successfully employ a model that is trained on a much larger cohort, and fine-tune it and still obtain significant improvements in performance. In our experiment, by using a pre-trained model that was trained on a large cohort of 2850 patients, to train on the Smart watch PPG dataset which had only 97 patients, we have showed an improved performance compared to training the model from scratch.

\section{Discussion and Conclusion}

The major finding of this study is that combining spectral representation of cardiac pulsatile recordings with traditional indices of heart rhythm irregularity in a deep neural network framework results in better AF classification. This approach facilitates transferring of learned model parameters across recording modalities such as ECG and PPG, thus enabling accurate AF classification in settings with limited access to large patient cohorts for model training purposes. Moreover, our results indicate that the hierarchical architecture of a deep neural network with one or more image-based feature extraction layers, a sequential layer capable of passing temporal information, and an  attention mechanism allows for accurate classification of paroxysmal AF. 

A key property of deep neural networks that contribute to their ability to generalize well is  learning of good representations of the data \cite{bengio2013representation}. Here the notion of \emph{good} is with respect to robustness to noise and invariance to factors that may contribute to irregularity of heart rhythms but may not be relevant to AF prediction. In the proposed method, this invariance is achieved through a combination of time-frequency representation, translation-invariance of CNN (due to the convolution operation and max pooling), and the attention mechanism that is capable of detecting important features independent of their actual position within a sequence of windows. For instance, having a higher heart rate shifts the position of the dominant frequency in the time-frequency plane representation of an ECG waveform, but the translation-invariance of the CNN makes the overall architecture robust to differences in baseline heart rates. Similarly, the time-frequency representation of a PPG waveforms exhibits dominant frequencies at the frequency of heart rate and the AF signature appears as an irregular distribution of power around this dominant frequency, and the CNN can effectively learn these irregular spectral patterns without getting confused by differences in heart rate across subjects. 

Various methods have been proposed in the literature for AF detection based on studying the dynamics of beat-to-beat heart rate time series extracted from ECG \cite{moody1983new,tateno2001automatic,dash2009automatic,lake2010accurate,linker2009long,colloca2013support,petrenas2012echo} and PPG \cite{nemati2016monitoring}. More recently, deep learning based methods have been shown to perform well in the context of AF detection. These deep learning based methods have harnessed the power of CNNs and RNNs \cite{shashikumar2017deep,rajpurkar2017cardiologist,clifford2017af} to capture the heart beat dynamics for detecting AF. All the previous works were designed for detection of AF over shorter segments of data (30 second to 90 second segments), and were not designed for the detection of paroxysmal AF which would require longer duration of data. In contrast, we consider ECG segments of 10 minute duration so as to capture paroxysmal AF which might occur as scattered episodes of AF in these segments. \citet{rajpurkar2017cardiologist} trained a 34 layer CNN to detect arrythmia from single lead ECG time series. In their proposed algorithm, the authors intentionally selected patients exhibiting abnormal rhythms to deal with class imbalance for both training and testing data, however this does not represent the actual prevalence of the rhythms in a real world setting, and will lead to optimistic results in the performance of the model. It is mentioned in the paper that the dataset consisted of 64,121 ECG records from 29,163 patients, with each ECG record being 30 seconds in duration, which is roughly around two 30 second windows for every patient. This method of choosing the ECG records was rather ambiguous. In the Physionet 2017 challenge, \citet{andreotti2017af} reimplemented the model proposed by \citet{rajpurkar2017cardiologist} and they were ranked 30th in the challenge. This showed that employing a vanilla neural network (in this case a vanilla CNN) on raw input data will not always provide the best performance in AF detection, and suggests that it would be more effective to use a combination of domain specific features and neural networks for AF detection.

Attention mechanisms in neural networks are inspired from the mechanism of visual attention found in humans. Attention mechanisms for neural networks have been well studied in the context of image recognition \cite{mnih2014recurrent,larochelle2010learning,denil2012learning}, and have only recently been applied to sequential prediction tasks. Some of the areas that they have been successfully applied to include neural machine translation \cite{bahdanau2014neural}, image caption generation \cite{xu2015show}, speech recognition \cite{chorowski2015attention}, and text summarization \cite{rush2015neural}. We employ a soft attention mechanism, which has the advantage that artificial segmentation of waveforms into positive and negative classes is not a pre-requirement of learning, but rather the algorithm is able to weight the different 30 seconds data segments to arrive at a final diagnosis of AF over the entire 10 minute window, in spite of occasional noise and non-AF rhythms within this window.  Moreover, the attention mechanism can provide a measure of the burden of AF, by quantifying the proportion of the 10 minute window where the attention weights are significantly positive. 

A major obstacle in application of machine learning techniques to clinical problems is the lack of acceptable clinical gold standards and scarcity of granular labels. Although, the PPG dataset utilized in this study was diligently annotated by multiple clinical adjudicators, the ECG dataset labels were obtained from automatic bedside Holter software, with a coarse overview by a single adjudicator for potential mislabeled cases of AF. Therefore, the ECG-based AF detection results provided in this work should be taken with some caution \cite{zhu2014crowd,zhu2015fusing,clifford2017af}. In fact, as demonstrated in \cref{table:afburden} re-defining the labels as a function of AF burden had a significant impact on the performance of the classifier (AUC of $0.94$ at $5\%$ or more versus AUC of $0.97$ at $75\%$ or more AF burden over a 10 minute segment). Although, our main goal for using the ECG dataset was to utilize the resulting model, trained on a relatively large patient cohort, in a transfer learning framework for prediction of AF in a much smaller cohort of patients with PPG recordings from a smart watch. Nevertheless, the proposed ECG-based algorithm, in concert with the visualization tool presented in \cref{fig:embeddings}, can be used in an active learning framework to further refine the labels by consulting with an expert panel of clinicians.

The paradigm presented here sets the stage for a new generation of real-time predictive analytics with broad applications across the growing numbers of devices with heart rate sensors. The Holter ECG dataset is a large scale, diverse training set with many different sub-categories of AF (fast, slow, paroxysmal, etc). As we proved in this study, such a database enables the production of more accurate algorithms in other devices in which such large datasets may not exist. Most notably, AF detection from PPG signal acquired from a wristband device (such as the Simband) is extremely challenging, given the low signal to noise ratio from such devices. Utilization of large datasets such as that from Holter ECG dataset with transfer learning help to enable algorithms with accuracies that may provide real-time AF detection from PPG wristband devices.

In summary, this study provides a well-motivated deep learning architecture for detection of paroxysmal AF, and demonstrates clinically acceptable AF detection accuracies across different recording modalities. To the best of our knowledge, this is the first study to consider the important problem of paroxysmal AF detection over long hours of recording using a deep learning framework with an attention mechanism. Our future work will be focused on prospective evaluation of the proposed technique in both intensive care unit patients and long-term recordings of patients with paroxysmal AF based on wearable technology.

\section{Acknowledgments and Funding}
Dr. Nemati is funded by the National Institutes of Health, award \#K01ES025445. The opinions or assertions contained herein are the private ones of the author/speaker and are not to be construed as official or reflecting the views of the National Institute of Health.

\bibliographystyle{ACM-Reference-Format}
\bibliography{rcnn}


\begin{thebibliography}{00}

%%% the Web address.

\ifx \showCODEN    \undefined \def \showCODEN     #1{\unskip}     \fi
\ifx \showDOI      \undefined \def \showDOI       #1{{\tt DOI:}\penalty0{#1}\ }
  \fi
\ifx \showISBNx    \undefined \def \showISBNx     #1{\unskip}     \fi
\ifx \showISBNxiii \undefined \def \showISBNxiii  #1{\unskip}     \fi
\ifx \showISSN     \undefined \def \showISSN      #1{\unskip}     \fi
\ifx \showLCCN     \undefined \def \showLCCN      #1{\unskip}     \fi
\ifx \shownote     \undefined \def \shownote      #1{#1}          \fi
\ifx \showarticletitle \undefined \def \showarticletitle #1{#1}   \fi
\ifx \showURL      \undefined \def \showURL       #1{#1}          \fi
% invisible to TeX
\providecommand\bibfield[2]{#2}
\providecommand\bibinfo[2]{#2}
\providecommand\natexlab[1]{#1}

\bibitem[\protect\citeauthoryear{Abadi, Barham, Chen, Chen, Davis, Dean, Devin,
  Ghemawat, Irving, Isard, Kudlur, Levenberg, Monga, Moore, Murray, Steiner,
  Tucker, Vasudevan, Warden, Wicke, Yu, and Zheng}{Abadi et~al\mbox{.}}{2016}]%
        {Tensorflow}
\bibfield{author}{\bibinfo{person}{Martin Abadi}, \bibinfo{person}{Paul
  Barham}, \bibinfo{person}{Jianmin Chen}, \bibinfo{person}{Zhifeng Chen},
  \bibinfo{person}{Andy Davis}, \bibinfo{person}{Jeffrey Dean},
  \bibinfo{person}{Matthieu Devin}, \bibinfo{person}{Sanjay Ghemawat},
  \bibinfo{person}{Geoffrey Irving}, \bibinfo{person}{Michael Isard},
  \bibinfo{person}{Manjunath Kudlur}, \bibinfo{person}{Josh Levenberg},
  \bibinfo{person}{Rajat Monga}, \bibinfo{person}{Sherry Moore},
  \bibinfo{person}{Derek~G. Murray}, \bibinfo{person}{Benoit Steiner},
  \bibinfo{person}{Paul Tucker}, \bibinfo{person}{Vijay Vasudevan},
  \bibinfo{person}{Pete Warden}, \bibinfo{person}{Martin Wicke},
  \bibinfo{person}{Yuan Yu}, {and} \bibinfo{person}{Xiaoqiang Zheng}.}
  \bibinfo{year}{2016}\natexlab{}.
\newblock \showarticletitle{TensorFlow: A system for Large-scale Machine
  Learning}. In \bibinfo{booktitle}{{\em 12th USENIX Symposium on Operating
  Systems Design and Implementation (OSDI 16)}}. \bibinfo{pages}{265--283}.
\newblock


\bibitem[\protect\citeauthoryear{Andreotti, Carr, Pimentel, Mahdi, and
  De~Vos}{Andreotti et~al\mbox{.}}{2017}]%
        {andreotti2017af}
\bibfield{author}{\bibinfo{person}{Fernando Andreotti}, \bibinfo{person}{Oliver
  Carr}, \bibinfo{person}{Marco A.~F. Pimentel}, \bibinfo{person}{Adam Mahdi},
  {and} \bibinfo{person}{Maarten De~Vos}.} \bibinfo{year}{2017}\natexlab{}.
\newblock \showarticletitle{Comparing Feature-Based Classifiers and
  Convolutional Neural Networks to Detect Arrhythmia from Short Segments of
  ECG}.
\newblock \bibinfo{journal}{{\em Computing in Cardiology\/}}
  \bibinfo{volume}{44} (\bibinfo{year}{2017}).
\newblock


\bibitem[\protect\citeauthoryear{Bahdanau, Cho, and Bengio}{Bahdanau
  et~al\mbox{.}}{2014}]%
        {bahdanau2014neural}
\bibfield{author}{\bibinfo{person}{Dzmitry Bahdanau},
  \bibinfo{person}{Kyunghyun Cho}, {and} \bibinfo{person}{Yoshua Bengio}.}
  \bibinfo{year}{2014}\natexlab{}.
\newblock \showarticletitle{Neural Machine Translation by Jointly Learning to
  Align and Translate}.
\newblock \bibinfo{journal}{{\em arXiv preprint arXiv:1409.0473\/}}
  (\bibinfo{year}{2014}).
\newblock


\bibitem[\protect\citeauthoryear{Ball, Carrington, McMurray, and Stewart}{Ball
  et~al\mbox{.}}{2013}]%
        {ball2013atrial}
\bibfield{author}{\bibinfo{person}{Jocasta Ball}, \bibinfo{person}{Melinda~J
  Carrington}, \bibinfo{person}{John~JV McMurray}, {and} \bibinfo{person}{Simon
  Stewart}.} \bibinfo{year}{2013}\natexlab{}.
\newblock \showarticletitle{Atrial fibrillation: Profile and Burden of an
  Evolving Epidemic in the 21st Century}.
\newblock \bibinfo{journal}{{\em International Journal of Cardiology\/}}
  \bibinfo{volume}{{167}, 5} (\bibinfo{year}{2013}),
  \bibinfo{pages}{1807--1824}.
\newblock


\bibitem[\protect\citeauthoryear{Behar, Oster, Li, and Clifford}{Behar
  et~al\mbox{.}}{2013}]%
        {behar2013ecg}
\bibfield{author}{\bibinfo{person}{Joachim Behar}, \bibinfo{person}{Julien
  Oster}, \bibinfo{person}{Qiao Li}, {and} \bibinfo{person}{Gari~D Clifford}.}
  \bibinfo{year}{2013}\natexlab{}.
\newblock \showarticletitle{ECG Signal Quality during Arrhythmia and its
  Application to Aalse Alarm Reduction}.
\newblock \bibinfo{journal}{{\em IEEE Transactions on Biomedical
  e=Engineering\/}} \bibinfo{volume}{{60}, 6} (\bibinfo{year}{2013}),
  \bibinfo{pages}{1660--1666}.
\newblock


\bibitem[\protect\citeauthoryear{Bengio, Courville, and Vincent}{Bengio
  et~al\mbox{.}}{2013}]%
        {bengio2013representation}
\bibfield{author}{\bibinfo{person}{Yoshua Bengio}, \bibinfo{person}{Aaron
  Courville}, {and} \bibinfo{person}{Pascal Vincent}.}
  \bibinfo{year}{2013}\natexlab{}.
\newblock \showarticletitle{Representation learning: A review and new
  perspectives}.
\newblock \bibinfo{journal}{{\em IEEE Transactions on Pattern Analysis and
  Machine Intelligence\/}} \bibinfo{volume}{{35}, 8} (\bibinfo{year}{2013}),
  \bibinfo{pages}{1798--1828}.
\newblock


\bibitem[\protect\citeauthoryear{Camm et~al\mbox{.}}{Camm
  et~al\mbox{.}}{2012}]%
        {camm20122012}
\bibfield{author}{\bibinfo{person}{A~John Camm} {and}
  \bibinfo{person}{others}.} \bibinfo{year}{2012}\natexlab{}.
\newblock \showarticletitle{2012 Focused Update of the ESC Guidelines for the
  Management of Atrial Fibrillation}.
\newblock \bibinfo{journal}{{\em European Heart Journal\/}}
  \bibinfo{volume}{{33}, 21} (\bibinfo{year}{2012}),
  \bibinfo{pages}{2719--2747}.
\newblock


\bibitem[\protect\citeauthoryear{Carlucci, Porzi, Caputo, Ricci, and
  Bul{\`o}}{Carlucci et~al\mbox{.}}{2017}]%
        {carlucci2017autodial}
\bibfield{author}{\bibinfo{person}{Fabio~Maria Carlucci},
  \bibinfo{person}{Lorenzo Porzi}, \bibinfo{person}{Barbara Caputo},
  \bibinfo{person}{Elisa Ricci}, {and} \bibinfo{person}{Samuel~Rota Bul{\`o}}.}
  \bibinfo{year}{2017}\natexlab{}.
\newblock \showarticletitle{Autodial: Automatic Domain Alignment Layers}. In
  \bibinfo{booktitle}{{\em International Conference on Computer Vision}}.
\newblock


\bibitem[\protect\citeauthoryear{Carrara, Carozzi, Moss, de~Pasquale, Cerutti,
  Ferrario, Lake, and Moorman}{Carrara et~al\mbox{.}}{2015}]%
        {carrara2015heart}
\bibfield{author}{\bibinfo{person}{Marta Carrara}, \bibinfo{person}{Luca
  Carozzi}, \bibinfo{person}{Travis~J Moss}, \bibinfo{person}{Marco de
  Pasquale}, \bibinfo{person}{Sergio Cerutti}, \bibinfo{person}{Manuela
  Ferrario}, \bibinfo{person}{Douglas~E Lake}, {and} \bibinfo{person}{J~Randall
  Moorman}.} \bibinfo{year}{2015}\natexlab{}.
\newblock \showarticletitle{Heart Rate Dynamics Distinguish Among Atrial
  Fibrillation, Normal Sinus Rhythm and Sinus Rhythm with Frequent Ectopy}.
\newblock \bibinfo{journal}{{\em Physiological Measurement\/}}
  \bibinfo{volume}{{36}, 9} (\bibinfo{year}{2015}), \bibinfo{pages}{1873}.
\newblock


\bibitem[\protect\citeauthoryear{Chorowski, Bahdanau, Serdyuk, Cho, and
  Bengio}{Chorowski et~al\mbox{.}}{2015}]%
        {chorowski2015attention}
\bibfield{author}{\bibinfo{person}{Jan~K Chorowski}, \bibinfo{person}{Dzmitry
  Bahdanau}, \bibinfo{person}{Dmitriy Serdyuk}, \bibinfo{person}{Kyunghyun
  Cho}, {and} \bibinfo{person}{Yoshua Bengio}.}
  \bibinfo{year}{2015}\natexlab{}.
\newblock \showarticletitle{Attention-based Models for Speech Recognition}. In
  \bibinfo{booktitle}{{\em Advances in Neural Information Processing Systems}}.
  \bibinfo{pages}{577--585}.
\newblock


\bibitem[\protect\citeauthoryear{Clifford, Liu, Moody, Lehman, Silva, Li,
  Johnson, and Mark}{Clifford et~al\mbox{.}}{2017a}]%
        {clifford2017af}
\bibfield{author}{\bibinfo{person}{Gari Clifford}, \bibinfo{person}{Chengyu
  Liu}, \bibinfo{person}{Benjamin Moody}, \bibinfo{person}{L Lehman},
  \bibinfo{person}{Ikaro Silva}, \bibinfo{person}{Qiao Li}, \bibinfo{person}{A
  Johnson}, {and} \bibinfo{person}{R Mark}.} \bibinfo{year}{2017}\natexlab{a}.
\newblock \showarticletitle{AF Classification from a Short Single Lead ECG
  Recording: The Physionet Computing in Cardiology Challenge 2017}.
\newblock \bibinfo{journal}{{\em Computing in Cardiology\/}}
  \bibinfo{volume}{44} (\bibinfo{year}{2017}).
\newblock


\bibitem[\protect\citeauthoryear{Clifford, Liu, Moody, Lehman, Silva, Li,
  Johnson, and Mark}{Clifford et~al\mbox{.}}{2017b}]%
        {Physionet2017Site}
\bibfield{author}{\bibinfo{person}{Gari Clifford}, \bibinfo{person}{Chengyu
  Liu}, \bibinfo{person}{Benjamin Moody}, \bibinfo{person}{L Lehman},
  \bibinfo{person}{Ikaro Silva}, \bibinfo{person}{Qiao Li}, \bibinfo{person}{A
  Johnson}, {and} \bibinfo{person}{R Mark}.} \bibinfo{year}{2017}\natexlab{b}.
\newblock \showarticletitle{Physionet Challenge 2017}.
\newblock \bibinfo{journal}{{\em https://physionet.org/challenge/2017/\/}}
  (\bibinfo{year}{2017}).
\newblock


\bibitem[\protect\citeauthoryear{Colloca, Johnson, Mainardi, and
  Clifford}{Colloca et~al\mbox{.}}{2013}]%
        {colloca2013support}
\bibfield{author}{\bibinfo{person}{Roberta Colloca},
  \bibinfo{person}{Alistair~EW Johnson}, \bibinfo{person}{Luca Mainardi}, {and}
  \bibinfo{person}{Gari~D Clifford}.} \bibinfo{year}{2013}\natexlab{}.
\newblock \showarticletitle{A Support Vector Machine Approach for Reliable
  Detection of Atrial Fibrillation Events}. In \bibinfo{booktitle}{{\em
  Computing in Cardiology Conference}}. IEEE, \bibinfo{pages}{1047--1050}.
\newblock


\bibitem[\protect\citeauthoryear{Cook and Ridker}{Cook and Ridker}{2009}]%
        {cook2009use}
\bibfield{author}{\bibinfo{person}{Nancy~R Cook} {and} \bibinfo{person}{Paul~M
  Ridker}.} \bibinfo{year}{2009}\natexlab{}.
\newblock \showarticletitle{The Use and Magnitude of Reclassification Measures
  for Individual Predictors of Global Cardiovascular Risk}.
\newblock \bibinfo{journal}{{\em Annals of Internal Medicine\/}}
  \bibinfo{volume}{{150}, 11} (\bibinfo{year}{2009}), \bibinfo{pages}{795}.
\newblock


\bibitem[\protect\citeauthoryear{Dash, Chon, Lu, and Raeder}{Dash
  et~al\mbox{.}}{2009}]%
        {dash2009automatic}
\bibfield{author}{\bibinfo{person}{S Dash}, \bibinfo{person}{KH Chon},
  \bibinfo{person}{S Lu}, {and} \bibinfo{person}{EA Raeder}.}
  \bibinfo{year}{2009}\natexlab{}.
\newblock \showarticletitle{Automatic Real Time Detection of Atrial
  Fibrillation}.
\newblock \bibinfo{journal}{{\em Annals of Biomedical Engineering\/}}
  \bibinfo{volume}{{37}, 9} (\bibinfo{year}{2009}),
  \bibinfo{pages}{1701--1709}.
\newblock


\bibitem[\protect\citeauthoryear{Daubechies}{Daubechies}{1990}]%
        {daubechies1990wavelet}
\bibfield{author}{\bibinfo{person}{Ingrid Daubechies}.}
  \bibinfo{year}{1990}\natexlab{}.
\newblock \showarticletitle{The Wavelet Transform, Time-frequency Localization
  and Signal Analysis}.
\newblock \bibinfo{journal}{{\em IEEE Transactions on Information Theory\/}}
  \bibinfo{volume}{{36}, 5} (\bibinfo{year}{1990}), \bibinfo{pages}{961--1005}.
\newblock


\bibitem[\protect\citeauthoryear{Denil, Bazzani, Larochelle, and
  de~Freitas}{Denil et~al\mbox{.}}{2012}]%
        {denil2012learning}
\bibfield{author}{\bibinfo{person}{Misha Denil}, \bibinfo{person}{Loris
  Bazzani}, \bibinfo{person}{Hugo Larochelle}, {and} \bibinfo{person}{Nando de
  Freitas}.} \bibinfo{year}{2012}\natexlab{}.
\newblock \showarticletitle{Learning where to attend with Deep Architectures
  for Image Tracking}.
\newblock \bibinfo{journal}{{\em Neural Computation\/}} \bibinfo{volume}{{24},
  8} (\bibinfo{year}{2012}), \bibinfo{pages}{2151--2184}.
\newblock


\bibitem[\protect\citeauthoryear{Drew, Harris, Z{\`e}gre-Hemsey, Mammone,
  Schindler, Salas-Boni, Bai, Tinoco, Ding, and Hu}{Drew et~al\mbox{.}}{2014}]%
        {drew2014insights}
\bibfield{author}{\bibinfo{person}{Barbara~J Drew}, \bibinfo{person}{Patricia
  Harris}, \bibinfo{person}{Jessica~K Z{\`e}gre-Hemsey}, \bibinfo{person}{Tina
  Mammone}, \bibinfo{person}{Daniel Schindler}, \bibinfo{person}{Rebeca
  Salas-Boni}, \bibinfo{person}{Yong Bai}, \bibinfo{person}{Adelita Tinoco},
  \bibinfo{person}{Quan Ding}, {and} \bibinfo{person}{Xiao Hu}.}
  \bibinfo{year}{2014}\natexlab{}.
\newblock \showarticletitle{Insights into the Problem of Alarm Fatigue with
  Physiologic Monitor Devices: a Comprehensive Observational Study of
  Consecutive Intensive Care Unit Patients}.
\newblock \bibinfo{journal}{{\em PloS One\/}} \bibinfo{volume}{{9}, 10}
  (\bibinfo{year}{2014}), \bibinfo{pages}{e110274}.
\newblock


\bibitem[\protect\citeauthoryear{Fuster, Ryd{\'e}n, Cannom, Crijns, Curtis,
  Ellenbogen, Halperin, Le~Heuzey, Kay, Lowe, et~al\mbox{.}}{Fuster
  et~al\mbox{.}}{2006}]%
        {fuster2006acc}
\bibfield{author}{\bibinfo{person}{Valentin Fuster}, \bibinfo{person}{Lars~E
  Ryd{\'e}n}, \bibinfo{person}{David~S Cannom}, \bibinfo{person}{Harry~J
  Crijns}, \bibinfo{person}{Anne~B Curtis}, \bibinfo{person}{Kenneth~A
  Ellenbogen}, \bibinfo{person}{Jonathan~L Halperin},
  \bibinfo{person}{Jean-Yves Le~Heuzey}, \bibinfo{person}{G~Neal Kay},
  \bibinfo{person}{James~E Lowe}, {and} \bibinfo{person}{others}.}
  \bibinfo{year}{2006}\natexlab{}.
\newblock \showarticletitle{ACC/AHA/ESC 2006 Guidelines for the Management of
  Patients with Atrial Fibrillation: full text: a Report of the American
  College of Cardiology/American Heart Association Task Force on Practice
  Guidelines and the European Society of Cardiology Committee for Practice
  Guidelines (Writing Committee to Revise the 2001 guidelines for the
  management of patients with atrial fibrillation) developed in collaboration
  with the European Heart Rhythm Association and the Heart Rhythm Society}.
\newblock \bibinfo{journal}{{\em Europace\/}} \bibinfo{volume}{{8}, 9}
  (\bibinfo{year}{2006}), \bibinfo{pages}{651--745}.
\newblock


\bibitem[\protect\citeauthoryear{Ghassemi, Lehman, Snoek, and Nemati}{Ghassemi
  et~al\mbox{.}}{2014}]%
        {ghassemi2014global}
\bibfield{author}{\bibinfo{person}{Mohammad Ghassemi}, \bibinfo{person}{Li-wei
  Lehman}, \bibinfo{person}{Jasper Snoek}, {and} \bibinfo{person}{Shamim
  Nemati}.} \bibinfo{year}{2014}\natexlab{}.
\newblock \showarticletitle{Global Optimization Approaches for Parameter Tuning
  in Biomedical Signal Processing: A Focus on Multi-scale Entropy}. In
  \bibinfo{booktitle}{{\em Computing in Cardiology Conference, 2014}}. IEEE,
  \bibinfo{pages}{993--996}.
\newblock


\bibitem[\protect\citeauthoryear{Graves, Mohamed, and Hinton}{Graves
  et~al\mbox{.}}{2013}]%
        {graves2013speech}
\bibfield{author}{\bibinfo{person}{Alex Graves}, \bibinfo{person}{Abdel-rahman
  Mohamed}, {and} \bibinfo{person}{Geoffrey Hinton}.}
  \bibinfo{year}{2013}\natexlab{}.
\newblock \showarticletitle{Speech Recognition with Deep Recurrent Neural
  Networks}. In \bibinfo{booktitle}{{\em Acoustics, Speech and Signal
  Processing (ICASSP), 2013 IEEE International Conference on}}. IEEE,
  \bibinfo{pages}{6645--6649}.
\newblock


\bibitem[\protect\citeauthoryear{Grinsted, Moore, and Jevrejeva}{Grinsted
  et~al\mbox{.}}{2004}]%
        {grinsted2004application}
\bibfield{author}{\bibinfo{person}{Aslak Grinsted}, \bibinfo{person}{John~C
  Moore}, {and} \bibinfo{person}{Svetlana Jevrejeva}.}
  \bibinfo{year}{2004}\natexlab{}.
\newblock \showarticletitle{Application of the Cross Wavelet Transform and
  Wavelet Coherence to Geophysical Time Series}.
\newblock \bibinfo{journal}{{\em Nonlinear Processes in Geophysics\/}}
  \bibinfo{volume}{{11}, 5/6} (\bibinfo{year}{2004}),
  \bibinfo{pages}{561--566}.
\newblock


\bibitem[\protect\citeauthoryear{He, Zhang, Ren, and Sun}{He
  et~al\mbox{.}}{2016}]%
        {he2016deep}
\bibfield{author}{\bibinfo{person}{Kaiming He}, \bibinfo{person}{Xiangyu
  Zhang}, \bibinfo{person}{Shaoqing Ren}, {and} \bibinfo{person}{Jian Sun}.}
  \bibinfo{year}{2016}\natexlab{}.
\newblock \showarticletitle{Deep Residual Learning for Image Recognition}. In
  \bibinfo{booktitle}{{\em Proceedings of the IEEE Conference on Computer
  Vision and Pattern Recognition}}. \bibinfo{pages}{770--778}.
\newblock


\bibitem[\protect\citeauthoryear{Hsu, Lv, and Kira}{Hsu et~al\mbox{.}}{2017}]%
        {hsu2017learning}
\bibfield{author}{\bibinfo{person}{Yen-Chang Hsu}, \bibinfo{person}{Zhaoyang
  Lv}, {and} \bibinfo{person}{Zsolt Kira}.} \bibinfo{year}{2017}\natexlab{}.
\newblock \showarticletitle{Learning to Cluster in Order to Transfer Across
  Domains and Tasks}.
\newblock \bibinfo{journal}{{\em arXiv preprint arXiv:1711.10125\/}}
  (\bibinfo{year}{2017}).
\newblock


\bibitem[\protect\citeauthoryear{Johnson, Behar, Andreotti, Clifford, and
  Oster}{Johnson et~al\mbox{.}}{2014}]%
        {johnson2014r}
\bibfield{author}{\bibinfo{person}{Alistair~EW Johnson},
  \bibinfo{person}{Joachim Behar}, \bibinfo{person}{Fernando Andreotti},
  \bibinfo{person}{Gari~D Clifford}, {and} \bibinfo{person}{Julien Oster}.}
  \bibinfo{year}{2014}\natexlab{}.
\newblock \showarticletitle{R-peak Estimation using Multimodal Lead Switching}.
  In \bibinfo{booktitle}{{\em Computing in Cardiology Conference}}. IEEE,
  \bibinfo{pages}{281--284}.
\newblock


\bibitem[\protect\citeauthoryear{Johnson, Behar, Andreotti, Clifford, and
  Oster}{Johnson et~al\mbox{.}}{2015}]%
        {johnson2015multimodal}
\bibfield{author}{\bibinfo{person}{Alistair~EW Johnson},
  \bibinfo{person}{Joachim Behar}, \bibinfo{person}{Fernando Andreotti},
  \bibinfo{person}{Gari~D Clifford}, {and} \bibinfo{person}{Julien Oster}.}
  \bibinfo{year}{2015}\natexlab{}.
\newblock \showarticletitle{Multimodal Heart Beat Detection using Signal
  Quality Indices}.
\newblock \bibinfo{journal}{{\em Physiological Measurement\/}}
  \bibinfo{volume}{{36}, 8} (\bibinfo{year}{2015}), \bibinfo{pages}{1665}.
\newblock


\bibitem[\protect\citeauthoryear{Krizhevsky, Sutskever, and Hinton}{Krizhevsky
  et~al\mbox{.}}{2012}]%
        {krizhevsky2012imagenet}
\bibfield{author}{\bibinfo{person}{Alex Krizhevsky}, \bibinfo{person}{Ilya
  Sutskever}, {and} \bibinfo{person}{Geoffrey~E Hinton}.}
  \bibinfo{year}{2012}\natexlab{}.
\newblock \showarticletitle{Imagenet Classification with Deep Convolutional
  Neural Networks}. In \bibinfo{booktitle}{{\em Advances in Neural Information
  Processing Systems}}. \bibinfo{pages}{1097--1105}.
\newblock


\bibitem[\protect\citeauthoryear{Lake and Moorman}{Lake and Moorman}{2010}]%
        {lake2010accurate}
\bibfield{author}{\bibinfo{person}{Douglas~E Lake} {and}
  \bibinfo{person}{J~Randall Moorman}.} \bibinfo{year}{2010}\natexlab{}.
\newblock \showarticletitle{Accurate Estimation of Entropy in Very Short
  Physiological Time Series: the Problem of Atrial Fibrillation Detection in
  Implanted Ventricular Devices}.
\newblock \bibinfo{journal}{{\em American Journal of Physiology-Heart and
  Circulatory Physiology\/}} \bibinfo{volume}{{300}, 1} (\bibinfo{year}{2010}),
  \bibinfo{pages}{H319--H325}.
\newblock


\bibitem[\protect\citeauthoryear{Larochelle and Hinton}{Larochelle and
  Hinton}{2010}]%
        {larochelle2010learning}
\bibfield{author}{\bibinfo{person}{Hugo Larochelle} {and}
  \bibinfo{person}{Geoffrey~E Hinton}.} \bibinfo{year}{2010}\natexlab{}.
\newblock \showarticletitle{Learning to Combine Foveal Glimpses with a
  third-order Boltzmann Machine}. In \bibinfo{booktitle}{{\em Advances in
  Neural Information Processing Systems}}. \bibinfo{pages}{1243--1251}.
\newblock


\bibitem[\protect\citeauthoryear{LeCun, Bengio, et~al\mbox{.}}{LeCun
  et~al\mbox{.}}{1995}]%
        {lecun1995convolutional}
\bibfield{author}{\bibinfo{person}{Yann LeCun}, \bibinfo{person}{Yoshua
  Bengio}, {and} \bibinfo{person}{others}.} \bibinfo{year}{1995}\natexlab{}.
\newblock \showarticletitle{Convolutional Networks for Images, Speech, and Time
  Series}.
\newblock \bibinfo{journal}{{\em The Handbook of Brain Theory and Neural
  Networks\/}} \bibinfo{volume}{{3361}, 10} (\bibinfo{year}{1995}),
  \bibinfo{pages}{1995}.
\newblock


\bibitem[\protect\citeauthoryear{LeCun, Bottou, Bengio, and Haffner}{LeCun
  et~al\mbox{.}}{1998}]%
        {lecun1998gradient}
\bibfield{author}{\bibinfo{person}{Yann LeCun}, \bibinfo{person}{L{\'e}on
  Bottou}, \bibinfo{person}{Yoshua Bengio}, {and} \bibinfo{person}{Patrick
  Haffner}.} \bibinfo{year}{1998}\natexlab{}.
\newblock \showarticletitle{Gradient-based Learning Applied to Document
  Recognition}.
\newblock \bibinfo{journal}{{\it Proc. IEEE}} \bibinfo{volume}{{86}, 11}
  (\bibinfo{year}{1998}), \bibinfo{pages}{2278--2324}.
\newblock


\bibitem[\protect\citeauthoryear{Lee~Giles, Kuhn, and Williams}{Lee~Giles
  et~al\mbox{.}}{1994}]%
        {lee1994dynamic}
\bibfield{author}{\bibinfo{person}{C Lee~Giles}, \bibinfo{person}{Gary~M Kuhn},
  {and} \bibinfo{person}{Ronald~J Williams}.} \bibinfo{year}{1994}\natexlab{}.
\newblock \showarticletitle{Dynamic Recurrent Neural Networks: Theory and
  Applications}.
\newblock \bibinfo{journal}{{\em IEEE Transactions on Neural Networks\/}}
  \bibinfo{volume}{{5}, 2} (\bibinfo{year}{1994}), \bibinfo{pages}{153--156}.
\newblock


\bibitem[\protect\citeauthoryear{Li, Mark, and Clifford}{Li
  et~al\mbox{.}}{2007}]%
        {li2007robust}
\bibfield{author}{\bibinfo{person}{Qiao Li}, \bibinfo{person}{Roger~G Mark},
  {and} \bibinfo{person}{Gari~D Clifford}.} \bibinfo{year}{2007}\natexlab{}.
\newblock \showarticletitle{Robust Heart Rate Estimation From Multiple
  Asynchronous Noisy Sources Using Signal Quality Indices and a Kalman Filter}.
\newblock \bibinfo{journal}{{\em Physiological Measurement\/}}
  \bibinfo{volume}{{29}, 1} (\bibinfo{year}{2007}), \bibinfo{pages}{15}.
\newblock


\bibitem[\protect\citeauthoryear{Linker}{Linker}{2009}]%
        {linker2009long}
\bibfield{author}{\bibinfo{person}{David~Thor Linker}.}
  \bibinfo{year}{2009}\natexlab{}.
\newblock \bibinfo{title}{Long-term Monitoring for Detection of Atrial
  Fibrillation}.
\newblock   (\bibinfo{date}{Dec.~8} \bibinfo{year}{2009}).
\newblock
\newblock
\shownote{US Patent 7,630,756.}


\bibitem[\protect\citeauthoryear{Lip and Hee}{Lip and Hee}{2001}]%
        {lip2001paroxysmal}
\bibfield{author}{\bibinfo{person}{Gregory~YH Lip} {and}
  \bibinfo{person}{FL~Li~Saw Hee}.} \bibinfo{year}{2001}\natexlab{}.
\newblock \showarticletitle{Paroxysmal Atrial Fibrillation}.
\newblock \bibinfo{journal}{{\em QJM: An International Journal of Medicine\/}}
  \bibinfo{volume}{{94}, 12} (\bibinfo{year}{2001}), \bibinfo{pages}{665--678}.
\newblock


\bibitem[\protect\citeauthoryear{Long, Wang, and Jordan}{Long
  et~al\mbox{.}}{2016}]%
        {long2016deep}
\bibfield{author}{\bibinfo{person}{Mingsheng Long}, \bibinfo{person}{Jianmin
  Wang}, {and} \bibinfo{person}{Michael~I Jordan}.}
  \bibinfo{year}{2016}\natexlab{}.
\newblock \showarticletitle{Deep Transfer Learning with Joint Adaptation
  Networks}.
\newblock \bibinfo{journal}{{\em arXiv preprint arXiv:1605.06636\/}}
  (\bibinfo{year}{2016}).
\newblock


\bibitem[\protect\citeauthoryear{Maaten and Hinton}{Maaten and Hinton}{2008}]%
        {maaten2008visualizing}
\bibfield{author}{\bibinfo{person}{Laurens van~der Maaten} {and}
  \bibinfo{person}{Geoffrey Hinton}.} \bibinfo{year}{2008}\natexlab{}.
\newblock \showarticletitle{Visualizing Data using t-SNE}.
\newblock \bibinfo{journal}{{\em Journal of Machine Learning Research\/}}
  \bibinfo{volume}{{9}, Nov} (\bibinfo{year}{2008}),
  \bibinfo{pages}{2579--2605}.
\newblock


\bibitem[\protect\citeauthoryear{Mart{\'\i}nez, Almeida, Olmos, Rocha, and
  Laguna}{Mart{\'\i}nez et~al\mbox{.}}{2004}]%
        {martinez2004wavelet}
\bibfield{author}{\bibinfo{person}{Juan~Pablo Mart{\'\i}nez},
  \bibinfo{person}{Rute Almeida}, \bibinfo{person}{Salvador Olmos},
  \bibinfo{person}{Ana~Paula Rocha}, {and} \bibinfo{person}{Pablo Laguna}.}
  \bibinfo{year}{2004}\natexlab{}.
\newblock \showarticletitle{A Wavelet-based ECG Delineator: Evaluation on
  Standard Databases}.
\newblock \bibinfo{journal}{{\em IEEE Transactions on Biomedical
  Engineering\/}} \bibinfo{volume}{{51}, 4} (\bibinfo{year}{2004}),
  \bibinfo{pages}{570--581}.
\newblock


\bibitem[\protect\citeauthoryear{MATLAB}{MATLAB}{2016}]%
        {MATLAB2016}
\bibfield{author}{\bibinfo{person}{MATLAB}.} \bibinfo{year}{2016}\natexlab{}.
\newblock \bibinfo{booktitle}{{\em version 9.1 (R2016b)}}.
\newblock The MathWorks Inc., Natick, Massachusetts.
\newblock


\bibitem[\protect\citeauthoryear{Mnih, Heess, Graves, et~al\mbox{.}}{Mnih
  et~al\mbox{.}}{2014}]%
        {mnih2014recurrent}
\bibfield{author}{\bibinfo{person}{Volodymyr Mnih}, \bibinfo{person}{Nicolas
  Heess}, \bibinfo{person}{Alex Graves}, {and} \bibinfo{person}{others}.}
  \bibinfo{year}{2014}\natexlab{}.
\newblock \showarticletitle{Recurrent Models of Visual Attention}. In
  \bibinfo{booktitle}{{\em Advances in Neural Information Processing Systems}}.
  \bibinfo{pages}{2204--2212}.
\newblock


\bibitem[\protect\citeauthoryear{Mnih, Kavukcuoglu, Silver, Rusu, Veness,
  Bellemare, Graves, Riedmiller, Fidjeland, Ostrovski, et~al\mbox{.}}{Mnih
  et~al\mbox{.}}{2015}]%
        {mnih2015human}
\bibfield{author}{\bibinfo{person}{Volodymyr Mnih}, \bibinfo{person}{Koray
  Kavukcuoglu}, \bibinfo{person}{David Silver}, \bibinfo{person}{Andrei~A
  Rusu}, \bibinfo{person}{Joel Veness}, \bibinfo{person}{Marc~G Bellemare},
  \bibinfo{person}{Alex Graves}, \bibinfo{person}{Martin Riedmiller},
  \bibinfo{person}{Andreas~K Fidjeland}, \bibinfo{person}{Georg Ostrovski},
  {and} \bibinfo{person}{others}.} \bibinfo{year}{2015}\natexlab{}.
\newblock \showarticletitle{Human-level Control through Deep Reinforcement
  Learning}.
\newblock \bibinfo{journal}{{\em Nature\/}} \bibinfo{volume}{{518}, 7540}
  (\bibinfo{year}{2015}), \bibinfo{pages}{529}.
\newblock


\bibitem[\protect\citeauthoryear{Moody}{Moody}{1983}]%
        {moody1983new}
\bibfield{author}{\bibinfo{person}{George Moody}.}
  \bibinfo{year}{1983}\natexlab{}.
\newblock \showarticletitle{A New Method for Detecting Atrial Fibrillation
  using RR Intervals}.
\newblock \bibinfo{journal}{{\em Computers in Cardiology\/}}
  (\bibinfo{year}{1983}), \bibinfo{pages}{227--230}.
\newblock


\bibitem[\protect\citeauthoryear{Nemati, Ghassemi, Ambai, Isakadze,
  Levantsevych, Shah, and Clifford}{Nemati et~al\mbox{.}}{2016}]%
        {nemati2016monitoring}
\bibfield{author}{\bibinfo{person}{Shamim Nemati}, \bibinfo{person}{Mohammad~M
  Ghassemi}, \bibinfo{person}{Vaidehi Ambai}, \bibinfo{person}{Nino Isakadze},
  \bibinfo{person}{Oleksiy Levantsevych}, \bibinfo{person}{Amit Shah}, {and}
  \bibinfo{person}{Gari~D Clifford}.} \bibinfo{year}{2016}\natexlab{}.
\newblock \showarticletitle{Monitoring and Detecting Atrial Fibrillation Using
  Wearable Technology}. In \bibinfo{booktitle}{{\em Engineering in Medicine and
  Biology Society (EMBC), 2016 IEEE 38th Annual International Conference of
  the}}. IEEE, \bibinfo{pages}{3394--3397}.
\newblock


\bibitem[\protect\citeauthoryear{Petrenas, Marozas, Sornmo, and
  Lukosevicius}{Petrenas et~al\mbox{.}}{2012}]%
        {petrenas2012echo}
\bibfield{author}{\bibinfo{person}{Andrius Petrenas}, \bibinfo{person}{Vaidotas
  Marozas}, \bibinfo{person}{Leif Sornmo}, {and} \bibinfo{person}{Arunas
  Lukosevicius}.} \bibinfo{year}{2012}\natexlab{}.
\newblock \showarticletitle{An Echo State Neural Network for QRST Cancellation
  during Atrial Fibrillation}.
\newblock \bibinfo{journal}{{\em IEEE Transactions on Biomedical
  Engineering\/}} \bibinfo{volume}{{59}, 10} (\bibinfo{year}{2012}),
  \bibinfo{pages}{2950--2957}.
\newblock


\bibitem[\protect\citeauthoryear{Rajpurkar, Hannun, Haghpanahi, Bourn, and
  Ng}{Rajpurkar et~al\mbox{.}}{2017}]%
        {rajpurkar2017cardiologist}
\bibfield{author}{\bibinfo{person}{Pranav Rajpurkar}, \bibinfo{person}{Awni~Y
  Hannun}, \bibinfo{person}{Masoumeh Haghpanahi}, \bibinfo{person}{Codie
  Bourn}, {and} \bibinfo{person}{Andrew~Y Ng}.}
  \bibinfo{year}{2017}\natexlab{}.
\newblock \showarticletitle{Cardiologist-level Arrhythmia Detection with
  Convolutional Neural Networks}.
\newblock \bibinfo{journal}{{\em arXiv preprint arXiv:1707.01836\/}}
  (\bibinfo{year}{2017}).
\newblock


\bibitem[\protect\citeauthoryear{Richman and Moorman}{Richman and
  Moorman}{2000}]%
        {richman2000physiological}
\bibfield{author}{\bibinfo{person}{Joshua~S Richman} {and}
  \bibinfo{person}{J~Randall Moorman}.} \bibinfo{year}{2000}\natexlab{}.
\newblock \showarticletitle{Physiological Time-series Analysis Using
  Approximate Entropy and Sample Entropy}.
\newblock \bibinfo{journal}{{\em American Journal of Physiology-Heart and
  Circulatory Physiology\/}} \bibinfo{volume}{{278}, 6} (\bibinfo{year}{2000}),
  \bibinfo{pages}{H2039--H2049}.
\newblock


\bibitem[\protect\citeauthoryear{Rush, Chopra, and Weston}{Rush
  et~al\mbox{.}}{2015}]%
        {rush2015neural}
\bibfield{author}{\bibinfo{person}{Alexander~M Rush}, \bibinfo{person}{Sumit
  Chopra}, {and} \bibinfo{person}{Jason Weston}.}
  \bibinfo{year}{2015}\natexlab{}.
\newblock \showarticletitle{A Neural Attention Model for Abstractive Sentence
  Summarization}.
\newblock \bibinfo{journal}{{\em arXiv preprint arXiv:1509.00685\/}}
  (\bibinfo{year}{2015}).
\newblock


\bibitem[\protect\citeauthoryear{Samsung}{Samsung}{2017}]%
        {simbandSite}
\bibfield{author}{\bibinfo{person}{Samsung}.} \bibinfo{year}{2017}\natexlab{}.
\newblock \showarticletitle{{Simband}'s Official Website}.
\newblock \bibinfo{journal}{{\em https://www.simband.io/\/}}
  (\bibinfo{year}{2017}).
\newblock


\bibitem[\protect\citeauthoryear{Schuster and Paliwal}{Schuster and
  Paliwal}{1997}]%
        {schuster1997bidirectional}
\bibfield{author}{\bibinfo{person}{Mike Schuster} {and}
  \bibinfo{person}{Kuldip~K Paliwal}.} \bibinfo{year}{1997}\natexlab{}.
\newblock \showarticletitle{Bidirectional Recurrent Neural Networks}.
\newblock \bibinfo{journal}{{\em IEEE Transactions on Signal Processing\/}}
  \bibinfo{volume}{{45}, 11} (\bibinfo{year}{1997}),
  \bibinfo{pages}{2673--2681}.
\newblock


\bibitem[\protect\citeauthoryear{Settles}{Settles}{2012}]%
        {settles2012active}
\bibfield{author}{\bibinfo{person}{Burr Settles}.}
  \bibinfo{year}{2012}\natexlab{}.
\newblock \showarticletitle{Active Learning}.
\newblock \bibinfo{journal}{{\em Synthesis Lectures on Artificial Intelligence
  and Machine Learning\/}} \bibinfo{volume}{{6}, 1} (\bibinfo{year}{2012}),
  \bibinfo{pages}{1--114}.
\newblock


\bibitem[\protect\citeauthoryear{Shashikumar, Shah, Li, Clifford, and
  Nemati}{Shashikumar et~al\mbox{.}}{2017}]%
        {shashikumar2017deep}
\bibfield{author}{\bibinfo{person}{Supreeth~Prajwal Shashikumar},
  \bibinfo{person}{Amit~J Shah}, \bibinfo{person}{Qiao Li},
  \bibinfo{person}{Gari~D Clifford}, {and} \bibinfo{person}{Shamim Nemati}.}
  \bibinfo{year}{2017}\natexlab{}.
\newblock \showarticletitle{A Deep Learning Approach to Monitoring and
  Detecting Atrial Fibrillation Using Wearable Technology}. In
  \bibinfo{booktitle}{{\em Biomedical \& Health Informatics (BHI), 2017 IEEE
  EMBS International Conference on}}. IEEE, \bibinfo{pages}{141--144}.
\newblock


\bibitem[\protect\citeauthoryear{Silver, Huang, Maddison, Guez, Sifre, Van
  Den~Driessche, Schrittwieser, Antonoglou, Panneershelvam, Lanctot,
  et~al\mbox{.}}{Silver et~al\mbox{.}}{2016}]%
        {silver2016mastering}
\bibfield{author}{\bibinfo{person}{David Silver}, \bibinfo{person}{Aja Huang},
  \bibinfo{person}{Chris~J Maddison}, \bibinfo{person}{Arthur Guez},
  \bibinfo{person}{Laurent Sifre}, \bibinfo{person}{George Van Den~Driessche},
  \bibinfo{person}{Julian Schrittwieser}, \bibinfo{person}{Ioannis Antonoglou},
  \bibinfo{person}{Veda Panneershelvam}, \bibinfo{person}{Marc Lanctot}, {and}
  \bibinfo{person}{others}.} \bibinfo{year}{2016}\natexlab{}.
\newblock \showarticletitle{Mastering the Game of Go with Deep Neural Networks
  and Tree Search}.
\newblock \bibinfo{journal}{{\em Nature\/}} \bibinfo{volume}{{529}, 7587}
  (\bibinfo{year}{2016}), \bibinfo{pages}{484--489}.
\newblock


\bibitem[\protect\citeauthoryear{Stewart, Hart, Hole, and McMurray}{Stewart
  et~al\mbox{.}}{2002}]%
        {stewart2002population}
\bibfield{author}{\bibinfo{person}{Simon Stewart}, \bibinfo{person}{Carole~L
  Hart}, \bibinfo{person}{David~J Hole}, {and} \bibinfo{person}{John~JV
  McMurray}.} \bibinfo{year}{2002}\natexlab{}.
\newblock \showarticletitle{A Population-based Study of the Long-term Risks
  Associated with Atrial Fibrillation: 20-year Follow-up of the Renfrew/Paisley
  Study}.
\newblock \bibinfo{journal}{{\em The American Journal of Medicine\/}}
  \bibinfo{volume}{{113}, 5} (\bibinfo{year}{2002}), \bibinfo{pages}{359--364}.
\newblock


\bibitem[\protect\citeauthoryear{Stridh and Sornmo}{Stridh and Sornmo}{2001}]%
        {stridh2001spatiotemporal}
\bibfield{author}{\bibinfo{person}{Martin Stridh} {and} \bibinfo{person}{L
  Sornmo}.} \bibinfo{year}{2001}\natexlab{}.
\newblock \showarticletitle{Spatiotemporal QRST Cancellation Techniques for
  Analysis of Atrial Fibrillation}.
\newblock \bibinfo{journal}{{\em IEEE Transactions on Biomedical
  Engineering\/}} \bibinfo{volume}{{48}, 1} (\bibinfo{year}{2001}),
  \bibinfo{pages}{105--111}.
\newblock


\bibitem[\protect\citeauthoryear{Sun, Feng, and Saenko}{Sun
  et~al\mbox{.}}{2016}]%
        {sun2016return}
\bibfield{author}{\bibinfo{person}{Baochen Sun}, \bibinfo{person}{Jiashi Feng},
  {and} \bibinfo{person}{Kate Saenko}.} \bibinfo{year}{2016}\natexlab{}.
\newblock \showarticletitle{Return of Frustratingly Easy Domain Adaptation.}.
  In \bibinfo{booktitle}{{\em AAAI}}, \bibinfo{volume}{Vol.~6}.
  \bibinfo{pages}{8}.
\newblock


\bibitem[\protect\citeauthoryear{Tateno and Glass}{Tateno and Glass}{2001}]%
        {tateno2001automatic}
\bibfield{author}{\bibinfo{person}{K Tateno} {and} \bibinfo{person}{L Glass}.}
  \bibinfo{year}{2001}\natexlab{}.
\newblock \showarticletitle{Automatic Detection of Atrial Fibrillation Using
  the Coefficient of Variation and Density Histograms of RR and $\Delta$RR
  Intervals}.
\newblock \bibinfo{journal}{{\em Medical and Biological Engineering and
  Computing\/}} \bibinfo{volume}{{39}, 6} (\bibinfo{year}{2001}),
  \bibinfo{pages}{664--671}.
\newblock


\bibitem[\protect\citeauthoryear{Tieleman and Hinton}{Tieleman and
  Hinton}{2012}]%
        {tieleman2012lecture}
\bibfield{author}{\bibinfo{person}{Tijmen Tieleman} {and}
  \bibinfo{person}{Geoffrey Hinton}.} \bibinfo{year}{2012}\natexlab{}.
\newblock \showarticletitle{Lecture 6.5-rmsprop: Divide the gradient by a
  running average of its recent magnitude}.
\newblock \bibinfo{journal}{{\em COURSERA: Neural networks for Machine
  Learning\/}} \bibinfo{volume}{{4}, 2} (\bibinfo{year}{2012}),
  \bibinfo{pages}{26--31}.
\newblock


\bibitem[\protect\citeauthoryear{Wong, Brooks, Leong, Roberts-Thomson, and
  Sanders}{Wong et~al\mbox{.}}{2012}]%
        {wong2012increasing}
\bibfield{author}{\bibinfo{person}{Christopher~X Wong},
  \bibinfo{person}{Anthony~G Brooks}, \bibinfo{person}{Darryl~P Leong},
  \bibinfo{person}{Kurt~C Roberts-Thomson}, {and} \bibinfo{person}{Prashanthan
  Sanders}.} \bibinfo{year}{2012}\natexlab{}.
\newblock \showarticletitle{The Increasing Burden of Atrial Fibrillation
  Compared with Heart Aailure and Myocardial Infarction: a 15-Year Study of All
  Hospitalizations in Australia}.
\newblock \bibinfo{journal}{{\em Archives of Internal Medicine\/}}
  \bibinfo{volume}{{172}, 9} (\bibinfo{year}{2012}), \bibinfo{pages}{739--741}.
\newblock


\bibitem[\protect\citeauthoryear{Wu, Schuster, Chen, Le, Norouzi, Macherey,
  Krikun, Cao, Gao, Macherey, et~al\mbox{.}}{Wu et~al\mbox{.}}{2016}]%
        {wu2016google}
\bibfield{author}{\bibinfo{person}{Yonghui Wu}, \bibinfo{person}{Mike
  Schuster}, \bibinfo{person}{Zhifeng Chen}, \bibinfo{person}{Quoc~V Le},
  \bibinfo{person}{Mohammad Norouzi}, \bibinfo{person}{Wolfgang Macherey},
  \bibinfo{person}{Maxim Krikun}, \bibinfo{person}{Yuan Cao},
  \bibinfo{person}{Qin Gao}, \bibinfo{person}{Klaus Macherey}, {and}
  \bibinfo{person}{others}.} \bibinfo{year}{2016}\natexlab{}.
\newblock \showarticletitle{Google's Neural Machine Translation System:
  Bridging the Gap between Human and Machine Translation}.
\newblock \bibinfo{journal}{{\em arXiv preprint arXiv:1609.08144\/}}
  (\bibinfo{year}{2016}).
\newblock


\bibitem[\protect\citeauthoryear{Xu, Ba, Kiros, Cho, Courville, Salakhudinov,
  Zemel, and Bengio}{Xu et~al\mbox{.}}{2015}]%
        {xu2015show}
\bibfield{author}{\bibinfo{person}{Kelvin Xu}, \bibinfo{person}{Jimmy Ba},
  \bibinfo{person}{Ryan Kiros}, \bibinfo{person}{Kyunghyun Cho},
  \bibinfo{person}{Aaron Courville}, \bibinfo{person}{Ruslan Salakhudinov},
  \bibinfo{person}{Rich Zemel}, {and} \bibinfo{person}{Yoshua Bengio}.}
  \bibinfo{year}{2015}\natexlab{}.
\newblock \showarticletitle{Show, Attend and Tell: Neural Image Caption
  Generation with Visual Attention}. In \bibinfo{booktitle}{{\em International
  Conference on Machine Learning}}. \bibinfo{pages}{2048--2057}.
\newblock


\bibitem[\protect\citeauthoryear{Yang, Yang, Dyer, He, Smola, and Hovy}{Yang
  et~al\mbox{.}}{2016}]%
        {yang2016hierarchical}
\bibfield{author}{\bibinfo{person}{Zichao Yang}, \bibinfo{person}{Diyi Yang},
  \bibinfo{person}{Chris Dyer}, \bibinfo{person}{Xiaodong He},
  \bibinfo{person}{Alexander~J Smola}, {and} \bibinfo{person}{Eduard~H Hovy}.}
  \bibinfo{year}{2016}\natexlab{}.
\newblock \showarticletitle{Hierarchical Attention Networks for Document
  Classification.}. In \bibinfo{booktitle}{{\em HLT-NAACL}}.
  \bibinfo{pages}{1480--1489}.
\newblock


\bibitem[\protect\citeauthoryear{Zhou, Shi, Tian, Qi, Li, Hao, and Xu}{Zhou
  et~al\mbox{.}}{2016}]%
        {zhou2016attention}
\bibfield{author}{\bibinfo{person}{Peng Zhou}, \bibinfo{person}{Wei Shi},
  \bibinfo{person}{Jun Tian}, \bibinfo{person}{Zhenyu Qi},
  \bibinfo{person}{Bingchen Li}, \bibinfo{person}{Hongwei Hao}, {and}
  \bibinfo{person}{Bo Xu}.} \bibinfo{year}{2016}\natexlab{}.
\newblock \showarticletitle{Attention-based Bidirectional Long Short-term
  Memory Networks for Relation Classification}. In \bibinfo{booktitle}{{\em
  Proceedings of the 54th Annual Meeting of the Association for Computational
  Linguistics (Volume 2: Short Papers)}}, \bibinfo{volume}{Vol.~2}.
  \bibinfo{pages}{207--212}.
\newblock


\bibitem[\protect\citeauthoryear{Zhu, Dunkley, Behar, Clifton, and
  Clifford}{Zhu et~al\mbox{.}}{2015}]%
        {zhu2015fusing}
\bibfield{author}{\bibinfo{person}{Tingting Zhu}, \bibinfo{person}{Nic
  Dunkley}, \bibinfo{person}{Joachim Behar}, \bibinfo{person}{David~A Clifton},
  {and} \bibinfo{person}{Gari~D Clifford}.} \bibinfo{year}{2015}\natexlab{}.
\newblock \showarticletitle{Fusing continuous-valued medical labels using a
  Bayesian model}.
\newblock \bibinfo{journal}{{\em Annals of Biomedical Engineering\/}}
  \bibinfo{volume}{{43}, 12} (\bibinfo{year}{2015}),
  \bibinfo{pages}{2892--2902}.
\newblock


\bibitem[\protect\citeauthoryear{Zhu, Johnson, Behar, and Clifford}{Zhu
  et~al\mbox{.}}{2014}]%
        {zhu2014crowd}
\bibfield{author}{\bibinfo{person}{Tingting Zhu}, \bibinfo{person}{Alistair~EW
  Johnson}, \bibinfo{person}{Joachim Behar}, {and} \bibinfo{person}{Gari~D
  Clifford}.} \bibinfo{year}{2014}\natexlab{}.
\newblock \showarticletitle{Crowd-sourced annotation of ECG signals using
  contextual information}.
\newblock \bibinfo{journal}{{\em Annals of Biomedical Engineering\/}}
  \bibinfo{volume}{{42}, 4} (\bibinfo{year}{2014}), \bibinfo{pages}{871--884}.
\newblock


\end{thebibliography}

\end{document}